%% file: Antineutrino_Energy_Spectrum_Unfolding_Based_on_the_Daya_Bay_Measurement_and_Its_Application.tex
\documentclass[a4paper,10pt,twoside]{cpc-hepnp}
\usepackage[colorlinks,
            linkcolor=blue,
            anchorcolor=blue,
            citecolor=blue]{hyperref}
\usepackage{hyperref}
\hypersetup{
    colorlinks=true,
    linkcolor=blue,
    filecolor=gray,      
    urlcolor=blue,
    citecolor=blue,
}

\usepackage{multicol}
\usepackage{graphicx}
\usepackage{booktabs}
\usepackage{amssymb,bm,mathrsfs,bbm,amscd}
\usepackage[usenames]{xcolor} 
\usepackage[tbtags]{amsmath}
\usepackage{lastpage}
\usepackage{CJK}
\usepackage{lineno}

\newcommand{\UL}{$^{235}$U}
\newcommand{\UH}{$^{238}$U}
\newcommand{\PuL}{$^{239}$Pu}
\newcommand{\PuH}{$^{241}$Pu}
\newcommand{\nuebar}{$\overline{\nu}_{e}$}

\begin{document}
\begin{CJK*}{GBK}{song}

\fancyhead[c]{\small Chinese Physics C~~~xxx} \fancyfoot[C]{\small yyy-\thepage}

\footnotetext[0]{Received \today}

\title{Antineutrino Energy Spectrum Unfolding Based on the Daya Bay Measurement and Its Applications
\footnotetext[2]{This work was supported in part by the Ministry of Science and Technology of China, the U.S. Department of Energy, the Chinese Academy of Sciences, the CAS Center for Excellence in Particle Physics, the National Natural Science Foundation of China, the Guangdong provincial government, the Shenzhen municipal government, the China General Nuclear Power Group, the Research Grants Council of the Hong Kong Special Administrative Region of China, the Ministry of Education in Taiwan, the U.S. National Science Foundation, the Ministry of Education, Youth, and Sports of the Czech Republic, the Charles University Research Centre UNCE, the Joint Institute of Nuclear Research in Dubna, Russia, the National Commission of Scientific and Technological Research of Chile, We acknowledge Yellow River Engineering Consulting Co., Ltd., and China Railway 15th Bureau Group Co., Ltd., for building the underground laboratory. We are grateful for the ongoing cooperation from the China Guangdong Nuclear Power Group and China Light \& Power Company. }}

\begin{CJK}{UTF8}{gbsn}
\input{CPC.tex}
\end{CJK}

\begin{abstract}
The prediction of reactor antineutrino spectra will play a crucial role as reactor experiments enter the precision era. The positron energy spectrum of 3.5 million antineutrino inverse beta decay reactions observed by the Daya Bay experiment, in combination with the fission rates of fissile isotopes in the reactor, is used to extract the positron energy spectra resulting from the fission of specific isotopes. This information can be used to produce a precise, data-based prediction of the antineutrino energy spectrum in other reactor antineutrino experiments with different fission fractions than Daya Bay. The positron energy spectra are unfolded to obtain the antineutrino energy spectra by removing the contribution from detector response with the Wiener-SVD unfolding method. Consistent results are obtained with other unfolding methods. A technique to construct a data-based prediction of the reactor antineutrino energy spectrum is proposed and investigated. Given the reactor fission fractions, the technique can predict the energy spectrum to a 2\% precision. In addition, we illustrate how to perform a rigorous comparison between the unfolded antineutrino spectrum and a theoretical model prediction that avoids the input model bias of the unfolding method. 
\end{abstract}



\footnotetext[0]{\hspace*{-3mm}\raisebox{0.3ex}{$\scriptstyle\copyright$}2013
Chinese Physical Society and the Institute of High Energy Physics
of the Chinese Academy of Sciences and the Institute
of Modern Physics of the Chinese Academy of Sciences and IOP Publishing Ltd}%


\section{Introduction}  
Nuclear reactors are a powerful source of electron antineutrinos (\nuebar) and have played a significant role in neutrino physics, including the discovery of neutrinos~\cite{Cowan:1992xc}, the measurement of the neutrino mixing angle $\theta_{12}$ and the neutrino mass-squared splitting $\Delta m^2_{21}$~\cite{Eguchi:2002dm}, and the observation of the neutrino oscillation driven by $\theta_{13}$~\cite{An:2012eh,Ahn:2012nd,Abe:2011fz}. Many very short baseline experiments are producing exciting results in the search for sterile neutrinos and in precise measurements of the reactor antineutrino energy spectrum~\cite{Siyeon:2017tsg,Ashenfelter:2018iov,Almazan:2018wln,Ashenfelter:2018jrx,AlmazanMolina:2020jlh}. Looking forward, resolution of the neutrino mass ordering is the design goal of the JUNO reactor neutrino experiment~\cite{An:2015jdp} at a baseline of 53~km. 

Commercial pressurized water reactors produce a large number of \nuebar 's emitted from the beta decay chains of the fission products from four main isotopes \UL, \UH, \PuL, and \PuH, while other isotopes contribute less than 0.3\%. In general, about 2$\times 10^{20}$ antineutrinos per second are released per GW thermal power.
The understanding of the antineutrino spectra is a key issue for reactor antineutrino experiments. 
At present there are two methods to obtain the predicted antineutrino spectra from these four fission isotopes. 

The first method is the conversion method. It is based on the measured beta spectra from thermal-neutron induced fission of \UL, \PuL, and \PuH~performed at the ILL High Flux Reactor~\cite{Schreckenbach:1985ep,VonFeilitzsch:1982jw,Hahn:1989zr}. 
The electron spectra of fission isotopes were fitted with a set of virtual beta decay branches based on allowed beta decay transitions. 
In 2011, Mueller {\it et al.} and Huber re-evaluated the flux and spectra including an improved beta spectrum calculation~\cite{Mueller:2011nm,Huber:2011wv} (the Huber-Mueller model). 
The uncertainties were estimated from detailed studies of corrections to the allowed $\beta$-spectrum shape, the inversion errors based on synthetic data sets, and the reliability of nuclear structure data. 
The re-evaluated flux was found to be about 5\% higher than past measurements~\cite{Mention:2011rk}. The discrepancy is commonly referred to as the ``Reactor Antineutrino Anomaly'' (RAA). 
Precise measurements of the reactor antineutrino spectral shape indicated another discrepancy (spectral distortion) compared to the Huber-Mueller model prediction~\cite{An:2015nua,Ko:2016owz,Abe:2014bwa,Seo:2014xei}.
A number of different hypotheses have been proposed to explain this discrepancy, such as improper treatment of the shape corrections for forbidden transitions~\cite{Hayes:2013wra}. 

The second method is the summation method. The total \nuebar~energy spectra of all known fission decay channels are calculated based on the fission yields of the fission products, Q values and decay branching fractions in the nuclear data libraries. 
This method generally has unknown uncertainties because the correlation among different sets of nuclear data uncertainties 
have not been properly cataloged. 
In addition, this method suffers from unknown uncertainties due to missing or incomplete information in the nuclear data libraries. 
It was recently pointed out that the bias of the pandemonium effect (the inability to accurately measure the complicated beta spectra for nuclei when the energy available for beta decay is large~\cite{Hardy:1977suw}) on nuclear structure measurements also impacts antineutrino summation predictions~\cite{Fallot:2012jv}.
Including pandemonium-free data of the major contributors can decrease the antineutrino flux from all fission isotopes~\cite{Estienne:2019ujo}, thus improving agreement with the measured flux from the fuel evolution study at Daya Bay~\cite{An:2017osx}.
Nonetheless, a recent comprehensive analysis with updated reactor antineutrino flux models showed the rate anomaly and spectral anomaly still persist~\cite{Berryman:2019hme}. 

The above studies illustrate the difficulty of constructing a reliable model because of the complexity of beta decay theory and the imperfect information in nuclear data libraries. 
To minimize the uncertainty due to the prediction of the antineutrino spectrum, a relative measurement comparing antineutrino events at near and far site detectors is utilized in the experiments aiming at the measurement of $\theta_{13}$~\cite{An:2012eh,Ahn:2012nd,Abe:2011fz}. In addition, the relative measurement technique is also utilized in experiments with segmented detectors for short-baseline sterile neutrino searches~\cite{Ashenfelter:2018iov,Almazan:2018wln}. The antineutrino flux and spectrum with little influence from neutrino oscillation are well measured in the near-site detectors. The uncertainty of the total measured antineutrino flux and spectrum is better than that of model predictions for commercial reactors~\cite{An:2016srz,Adey:2019ywk}. 
These absolute spectrum measurements can be used as an additional resource for validating standard nuclear databases~\cite{Bernstein:2019nqq}.
Moreover, these measurements can also provide a reference spectrum for other reactor antineutrino experiments, especially with single unsegmented detectors. 
These measurements have been utilized in the studies of reactor antineutrino experiments via the inverse beta decay (IBD) reaction~\cite{Ko:2016owz,An:2015jdp}. These precise measurements can also be used as an input for reactor antineutrino experiments utilizing coherent elastic neutrino-nucleus scattering (CE$\nu$NS)~\cite{Aguilar-Arevalo:2019jlr, Bonet:2020awv}.

Reactor antineutrinos are measured via the IBD reaction at Daya Bay. The positron carries most of the antineutrino energy and forms a prompt signal in the detector. In Ref.~\cite{Adey:2019ywk}, the measured prompt energy spectrum based on 1958 days of data acquisition was provided in bins of 0.25~MeV. 
Moreover, the prompt energy spectra of \UL~and \PuL~were extracted for the first time based on the commercial reactor data. 
To provide a data-driven prediction for other experiments with different fission fractions than Daya Bay, the correlation of the total measured prompt energy spectrum and the extracted prompt energy spectra for specific fission isotopes is obtained(the term ``isotopic energy spectrum'' will be used to refer to an energy spectrum for a specific fission isotope in this manuscript). 
Furthermore, the corresponding antineutrino energy spectra are presented and utilized in this paper, by removing the contribution from detector response with the Wiener-SVD unfolding method~\cite{Tang:2017rob}. 
Finally, a new method is utilized to predict the antineutrino spectrum for other experiments at reactors with arbitrary fission fractions. 

This paper is organized as follows: Section~{\ref{Sec-DayaBayIntro}} introduces the antineutrino detectors, the previous studies, the event selection, and the detector response at Daya Bay. 
Section~{\ref{Sec-CorrelationAna}} provides the correlation of the total measurement and the extracted isotopic spectra. Section~{\ref{Sec-Unfolding}} presents the generic \nuebar~energy spectra of the IBD reaction. Application of generic \nuebar~energy spectra is explained in detail in Sec.~{\ref{Sec-Application}}. A short summary is given in Sec.~{\ref{Sec-Summary}}.

\section{Introduction to the Daya Bay experiment}
\label{Sec-DayaBayIntro}

The Daya Bay experiment studies the flux and spectrum of \nuebar's from six $2.9~\mathrm{GW}$ thermal power commercial pressurized-water reactor cores at the Daya Bay nuclear power complex. 
Eight identically designed antineutrino detectors (ADs) are distributed at two near-site experiment halls (EH1 and EH2) and one far-site experiment hall (EH3). The near site halls are used to monitor the \nuebar~flux and spectrum from the reactor cores (with flux-weighted baselines of 560~m and 600~m for EH1 and EH2, respectively), while the far-site experiment hall is used to measure the oscillated spectrum and flux deficit driven by $\theta_{13}$ (with flux-weighted baselines of 1640~m). 
Each AD consists of three nested cylindrical vessels. The inner acrylic vessel (IAV) contains 20 tons of 0.1\% gadolinium-loaded liquid scintillator (GdLS) and serves as the central \nuebar~target. The outer acrylic vessel surrounding the target creates a 42-cm thick pure liquid scintillator (LS) region to improve the collection of gammas escaping from the GdLS region to reduce the energy leakage. Scintillation light is detected by 192 8-inch PMTs (Hamamatsu R5912), which are positioned on the outermost cylinder of stainless steel. A mineral oil buffer shields the GdLS and LS regions from gamma rays from natural radioactivity in the PMTs. The experimental setup is described in detail in Refs.~\cite{Adey:2018zwh, An:2015qga}.

In a commercial reactor core, the chain reaction enabled by the neutrons produced through the fission of uranium and plutonium isotopes maintains the overall burn-up of the nuclear reactor fuel. The fissile isotopes \PuL~and \PuH~are readily produced by neutron capture on \UH~and ensuing reactions. The fraction of nuclear fissions attributed to a parent isotope, such as \PuL, is called the fission fraction. The fission fractions of different isotopes evolve with burn-up. Figure~\ref{Fig-FissionFraction} shows an example of the evolution of the fission fractions as a function of burn-up within a refueling cycle~\cite{An:2016srz}. The fission fraction of \UL~decreases with burn-up while the fission fractions of \PuL~and \PuH~increase. The fission fraction of \UH~is relatively small (7.6\%) and stable over time. Over 99.7\% of the $\bar\nu_e$'s produced from thousands of beta decays of fission daughters are due to the four isotopes \UL, \PuL, \UH~and \PuH.

\begin{center}
\includegraphics[width=0.5\columnwidth]{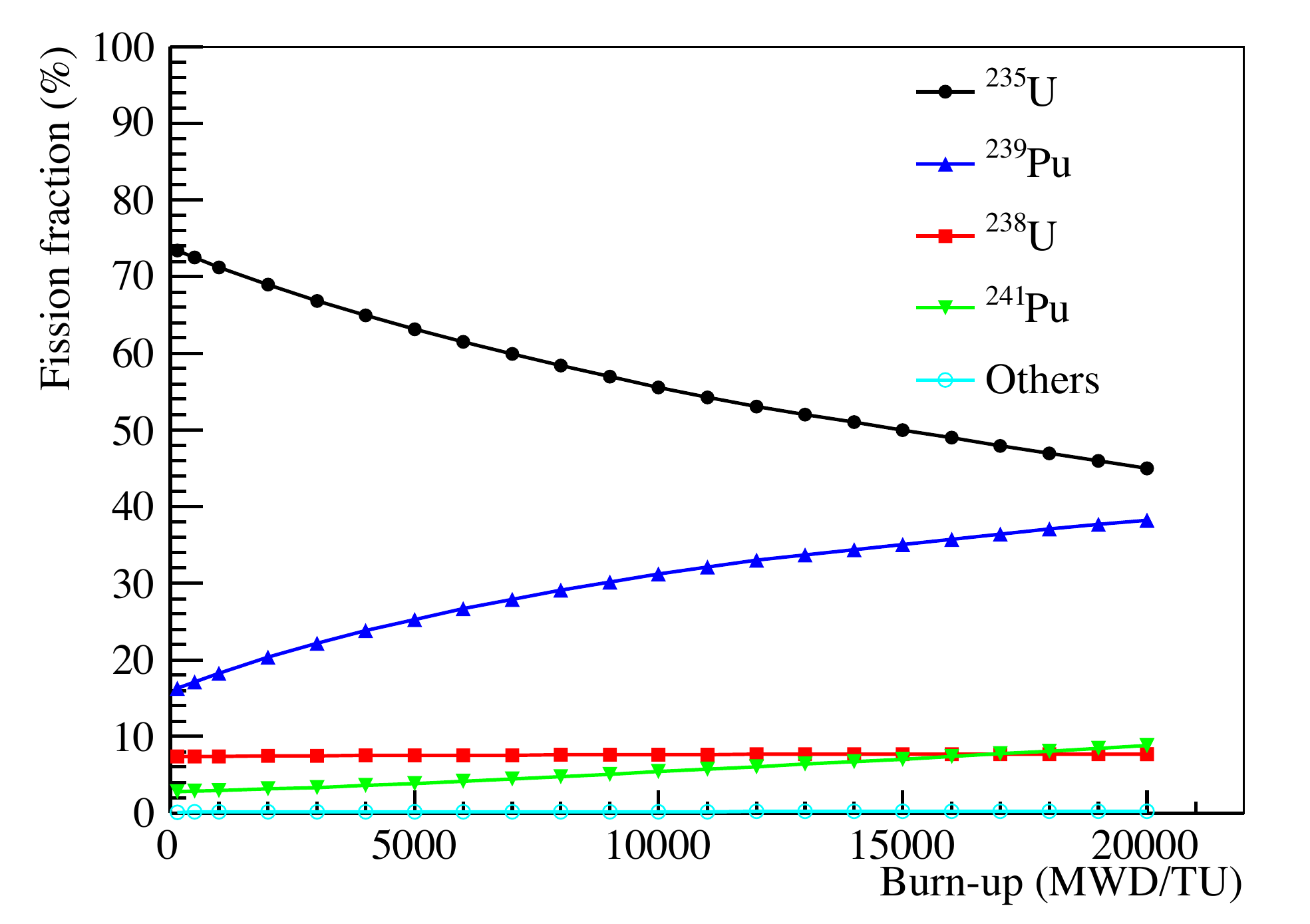}
\figcaption{\label{Fig-FissionFraction}  Fission fractions of isotopes as a function of burn-up from a simulation of a reactor core at the Daya Bay nuclear power plant. Other isotopes contribute less than 0.3\% in total. }
\end{center}

The Daya Bay experiment has measured the total flux and spectrum of reactor antineutrinos~\cite{An:2015nua, An:2016srz, Adey:2018qct}. Based on the evolution with respect to the fission fractions, the \nuebar~flux and spectra of two primary fission isotopes, \UL~and \PuL, were extracted~\cite{An:2017osx, Adey:2019ywk}. These results have contributed significantly to the understanding of the RAA. 

The first measurement of the antineutrino flux and the prompt energy spectrum with $\sim$0.3 million IBD events were reported in Ref.~\cite{An:2015nua}. A $\sim$5\% flux deficit, and a $\sim$10\% spectral distortion in the 4-6~MeV prompt energy region were found, relative to the Huber-Mueller model predictions. The details of this analysis were published in Ref.~\cite{An:2016srz} with an increased IBD sample of about 1.2~million events. Moreover, the antineutrino energy spectrum weighted by the IBD cross section was obtained. It has provided model-independent predictions for other reactor antineutrino experiments. 
To reduce the systematic uncertainties in the flux and spectrum measurements, the detection efficiency and the energy response were carefully determined~\cite{Adey:2018qct,Adey:2019zfo}. These efforts led to the improved flux measurement in Ref.~\cite{Adey:2018qct}, and the improved spectrum measurement in Ref.~\cite{Adey:2019ywk}.
Two novel measurements were performed by introducing the information about reactor fuel burning. The antineutrino fluxes of \UL~and \PuL~were extracted in Ref.~\cite{An:2017osx} using 2.2 million IBD candidates. In Ref.~\cite{An:2017osx}, a 7.8\% discrepancy between the measured and predicted \UL~antineutrino yield was found, indicating that \UL~is likely the primary contributor to the RAA.
In addition, the measurement at Daya Bay disfavored the hypotheses of equal deficits for all fission isotopes and a \PuL-only deficit as the reason for the discrepancy with the predicted flux at the 2.8$\sigma$ and 3.2$\sigma$ confidence levels, respectively.
The individual prompt energy spectra of \UL~and \PuL~were extracted for the first time in commercial reactors~\cite{Adey:2019ywk} using 3.5~million IBD candidates. 
The work in this paper uses the same data set as Ref.~\cite{Adey:2019ywk}.

This analysis follows the work in Ref.~\cite{Adey:2019ywk}, which makes use of 3.5 million inverse beta decay candidates from the four near-site ADs.
IBD candidates are selected following the same criteria as Ref.~\cite{An:2016ses}. 
After applying these cuts, the estimated signal and background rates, as well as the efficiencies of the muon veto, $\epsilon_{\mu}$, and multiplicity selection, $\epsilon_{\rm m}$, were determined~\cite{Adey:2018zwh}. Backgrounds remaining in the IBD samples from accidental coincidences, fast neutrons, cosmogenic $^8$He/$^9$Li production, AD-intrinsic alpha radioactivity, and AmC neutron calibration sources were estimated using a variety of techniques described in detail in a previous publication \cite{An:2012bu}. Additionally, spent nuclear fuel (SNF) present in the cooling pool adjacent to each reactor core contributes $\sim$0.3\% to the IBD rate \cite{An:2016srz}. The effect of the SNF was subtracted from the IBD spectrum. The relative uncertainty from SNF was reduced from 100\% \cite{An:2016ses} to 30\% \cite{Ma:2015lsv} using the spent fuel inventory history provided by the nuclear power plant. The relative rate of background is less than 2\% and contributes less than 0.15\% to the uncertainty on the IBD rate and less than 0.02\% to the uncertainty of IBD shape in the 1.5-6.0~MeV prompt energy region.  

The reconstructed positron energy relative to the true interaction energy is nonlinear due to ionization quenching and the Cherenkov light emission in the scintillator, and the underestimate of the charge of the PMT signals in the readout system. Previously, the energy nonlinearity model was constructed with 1\% uncertainty based on a semi-empirical analytic approach~\cite{An:2016ses}. 
In December 2015, a full Flash-ADC (FADC) readout system was installed at EH1 AD1, which recorded the PMT waveforms simultaneously with the previous front end electronics (FEE) systems. 
Based on a deconvolution method, the integral charge was extracted with a minimum bias based on the PMT waveform recorded by the FADC~\cite{Huang:2017abb}. The integral charge from the FADC was compared with the reconstructed charge from the FEE. The uncertainty on the electronics nonlinearity was reduced to 0.2\% based on an event-by-event comparison of the total charge from these two readout systems. In addition, the uncertainty in the visible energy from $\gamma$ rays was improved from 1\% to 0.5\% after a special calibration campaign in January 2017 that deployed $^{60}$Co sources with different enclosures to quantify the optical shadowing effect. Finally, the statistics of cosmogenic $^{12}$B candidates detected in four near-site ADs increased to $\sim$470,000 which were used to further refine the nonlinearity model. Based on these improvements, the uncertainty on the energy nonlinearity was improved to $<$0.5\% for prompt energy larger than 2~MeV. More details can be found in Refs.~\cite{Adey:2019zfo,Adey:2018zwh}.

Comparison of the mean reconstructed energy between antineutrino detectors was done with a variety of calibration references (neutron-capture on hydrogen and gadolinium, gammas from external $^{40}$K and $^{208}$Tl decays, and $\alpha$'s from $^{212}$Po, $^{215}$Po and $^{219}$Po decays). The variations were less than 0.2\%~\cite{An:2016ses}.

As mentioned above, the IAV contains the GdLS (\nuebar~target), but the IAV is non-scintillating material. Positrons and the annihilation $\gamma$-rays from IBD reactions around or inside the acrylic may lose energy invisibly in the acrylic, and this phenomenon leads to energy leakage and a slight distortion of the prompt energy spectrum. The relative uncertainty on the measured prompt energy spectrum from this effect (called the ``IAV effect'' in this manuscript) was estimated to be 4\% (0.1\%) below (above) 1.25~MeV by using simulation and it is assumed to be correlated among detectors~\cite{An:2016srz}. 

The energy resolution of the detectors was studied based on the measured reconstructed energy spectra of a variety of calibration sources deployed at the detector center, IBD and spallation neutrons, and alpha sources from radioactivity. The relative energy resolution of the ADs as a function of energy was modeled using the expression
\begin{align}
\frac{\sigma_E}{E_{\rm rec}}=\sqrt{a^2+\frac{b^2}{E_{\rm rec}}+\frac{c^2}{E_{\rm rec}^2}}, \label{Eq-EnergyResolution}
\end{align}
where $\sigma_E$ is the standard deviation of the reconstructed energy $E_{\rm rec}$, and the parameters $a$, $b$, and $c$ quantify the contribution to the resolution from detector energy nonuniformity, photoelectron statistics, and PMT dark noise, respectively. The best-fit parameters of the model are $a$ = 0.016, $b$ = 0.081~MeV$^{1/2}$ and $c$ = 0.026~MeV. 
The relative degradation of the energy resolution is less than 6\% after 6 years of operation ($\sim$8.5\% to $\sim$9.0\% at $E_{\rm rec}$=1~MeV), and the degradation has negligible affect on following analysis. 

The response matrix mapping the \nuebar~energy to the reconstructed energy was constructed using a full-detector simulation based on Geant4 \cite{Agostinelli:2002hh}. 
In this case, the response matrix included the IBD energy shift (the energy shift when \nuebar~energy is transferred to a positron and reconstructed as prompt energy), the IAV effect, and energy resolution. 
The reconstructed energy of IBD events is corrected for energy nonlinearity, which means the reconstructed energy of IBD events is the energy deposited in the LS. 
The response matrix was provided in the supplemental materials of Ref.~\cite{Adey:2019ywk}. For convenience, it is provided with the supplemental materials of this paper as well. 

\section{Total measurement and extracted isotopic prompt energy spectra}
\label{Sec-CorrelationAna}
The total measurement of the prompt energy spectrum in this paper is based on the data from the four near-site ADs. 
For each AD, the detected \nuebar 's are from 6 reactor cores with different baselines. 
Antineutrinos from each reactor come from four main isotopes with different relative contributions, hence the total prompt energy spectrum is the sum of the four isotopic energy spectra of prompt signal. 
To define the fractional contribution from different isotopes to the total number of IBD events in near-site ADs, the effective fission fraction of the total prompt energy spectrum is defined as:
\begin{align} 
f^{\rm iso}_{\rm eff}=\frac{\sum_{d=1}^{4}\sum_{r=1}^{6}N_r^f/L_{dr}^2\cdot f_r^{\rm iso}}{\sum_{d=1}^{4}\sum_{r=1}^{6}N_r^f/L_{dr}^2}.  \label{a0}
\end{align} 
Here $f_r^{\rm iso}$ is the fission fraction of $r$-th core for each isotope; $L_{dr}$ is the distance between the $d$-th detector and the $r$-th reactor core; $N_r^f$ is the predicted number of fissions for each isotope from the $r$-th reactor core, and it is calculated based on: 
\begin{align} 
N_r^f=\int \frac{W_r(t)}{\sum_{\rm iso}f_r^{\rm iso}(t)E_{\rm iso}}dt, 
\end{align} 
 where $t$ represents the detector data acquisition (DAQ) live time, $W_r$ represents the thermal power of $r$-th core, $E_{\rm iso}$ represents the mean energy release per fission for each isotope. The average effective fission fractions for \UL, \UH, \PuL~and \PuH~in near-site ADs for the analyzed data set are $f^{235}_{\rm eff}:f^{238}_{\rm eff}:f^{239}_{\rm eff}:f^{241}_{\rm eff}=0.564: 0.076: 0.304: 0.056$. 
The detailed description of the measurement as well as the comparison with the Huber-Mueller model can be found in Ref.~\cite{Adey:2019ywk}.

As the observed prompt energy spectrum evolves as a function of fission fractions, the isotopic prompt energy spectra from \UL~and \PuL~have been extracted at Daya Bay~\cite{Adey:2019ywk}. 
Daya Bay data are less sensitive to individual spectra of \UH~and \PuH~due to their smaller fission fractions than \UL~and \PuL. 
Constraints on individual spectra of \UH~and \PuH~are needed to obtain the dominant isotopic prompt energy spectra from \UL~and \PuL. 
The constraints on the prompt energy spectra of \UH~and \PuH~are given in Ref.~\cite{Adey:2019ywk}. 
Since the fission fraction of \PuH~evolves approximately proportional to \PuL~(the ratio is fitted to be 0.183), the spectra of \PuL~($s_{239}$) and \PuH~($s_{241}$) can be treated as one component, defined as $s_{\rm combo} = s_{239} + R^{\rm Pu} \times s_{241}$. Here $R^{\rm Pu}$ should be a constant to make $s_{\rm combo}$ an invariant and $R^{\rm Pu}$ is chosen to be 0.183 to have the least dependence on the input of the \PuH~spectrum, thus leading to the best precision of Pu combo spectrum. 
The residual contribution of \PuH~spectrum on the data was corrected when the fission fraction ratios of \PuH~to \PuL~deviate from 0.183.
The extracted isotopic (\UL~and Pu combo) prompt energy spectra of the IBD reaction were published in Ref.~\cite{Adey:2019ywk}, which provided uncertainties for both the flux and spectral shape. 

The isotopic energy spectra have larger uncertainties than the total observed prompt energy spectrum since the individual components extracted from the total one have dependence on extra information, which brings in additional uncertainties. 
The total prompt energy spectrum can provide a data-driven prediction for other reactor \nuebar~experiments with similar fission fractions. 
The isotopic prompt energy spectra can be used to correct effects from the differences in fission fractions between experiments. 
Since the total and isotopic prompt energy spectra are derived from the same data set, their uncertainties are correlated, which has to be taken into account when using them together. 
The total prompt energy spectrum, $s_{\rm total}$, is the sum of isotopic prompt energy spectra times the corresponding effective fission fractions:
\begin{align}
    s_{\rm total}=&f_{235}\cdot s_{235}+f_{239}\cdot s_{\rm combo}+f_{238}\cdot s_{238}\nonumber\\
    &+(f_{241}-0.183\times f_{239})\cdot s_{241}.  \label{Eq-totalspectrum}
\end{align}
The spectrum of Pu combo is used in this analysis because it has a smaller uncertainty than the total uncertainty of the extracted prompt energy spectrum of \PuL~and the correlation between the fission fractions of \PuL~and \PuH~in most commercial low-enriched uranium (LEU) nuclear power reactors are similar. 
The \PuL~spectrum can be extracted by subtracting prediction of the prompt energy spectrum based on the \PuH~model.

Based on the covariance matrix among the isotopic prompt energy spectra from the extraction algorithm \cite{Adey:2019ywk}, the total prompt energy spectrum and its correlation with the other two isotopic prompt energy spectra ($s_{235}$ and $s_{\rm combo}$) are obtained with Eq.~\ref{Eq-totalspectrum} through standard error propagation. 
The total and isotopic prompt energy spectra, as well as the covariance matrix are provided in the supplemental materials. 
Dominant components of the energy-dependent uncertainties for total and isotopic (\UL~and Pu combo) prompt energy spectra are shown in Fig.~\ref{Fig-PromptErrBudget}. Detector response uncertainty contains the contribution of the uncertainties from detection efficiency, the energy nonlinearity model, the energy scale difference between ADs, and the IAV effect. Model (\UH, \PuH) uncertainty represents uncertainties of the input \UH~and \PuH~model in the analysis to extract the isotopic (\UL~and Pu combo) prompt energy spectra. The uncertainties of the isotopic prompt energy spectra are dominated by statistical uncertainty and the uncertainty of the models (for \UH~and \PuH). 
Other uncertainties (e.g. background uncertainty) have small contribution to the total uncertainty and they are not shown in the plot. 
The calculated total measured prompt energy spectrum and its uncertainty in this analysis are consistent with the previous results \cite{Adey:2019ywk}. 
The uncertainty of the total measurement is dominated by the detector response uncertainty, especially the uncertainty of the detection efficiency (1.19\%). The primary not-fully-energy-correlated uncertainties for the total measurement are dominated by the uncertainty of the energy nonlinearity model. 
There is no contribution of the energy-dependent uncertainties from the input \UH~and \PuH~models to the total measurement since it is a direct measurement.

\begin{center}
\includegraphics[width=\columnwidth]{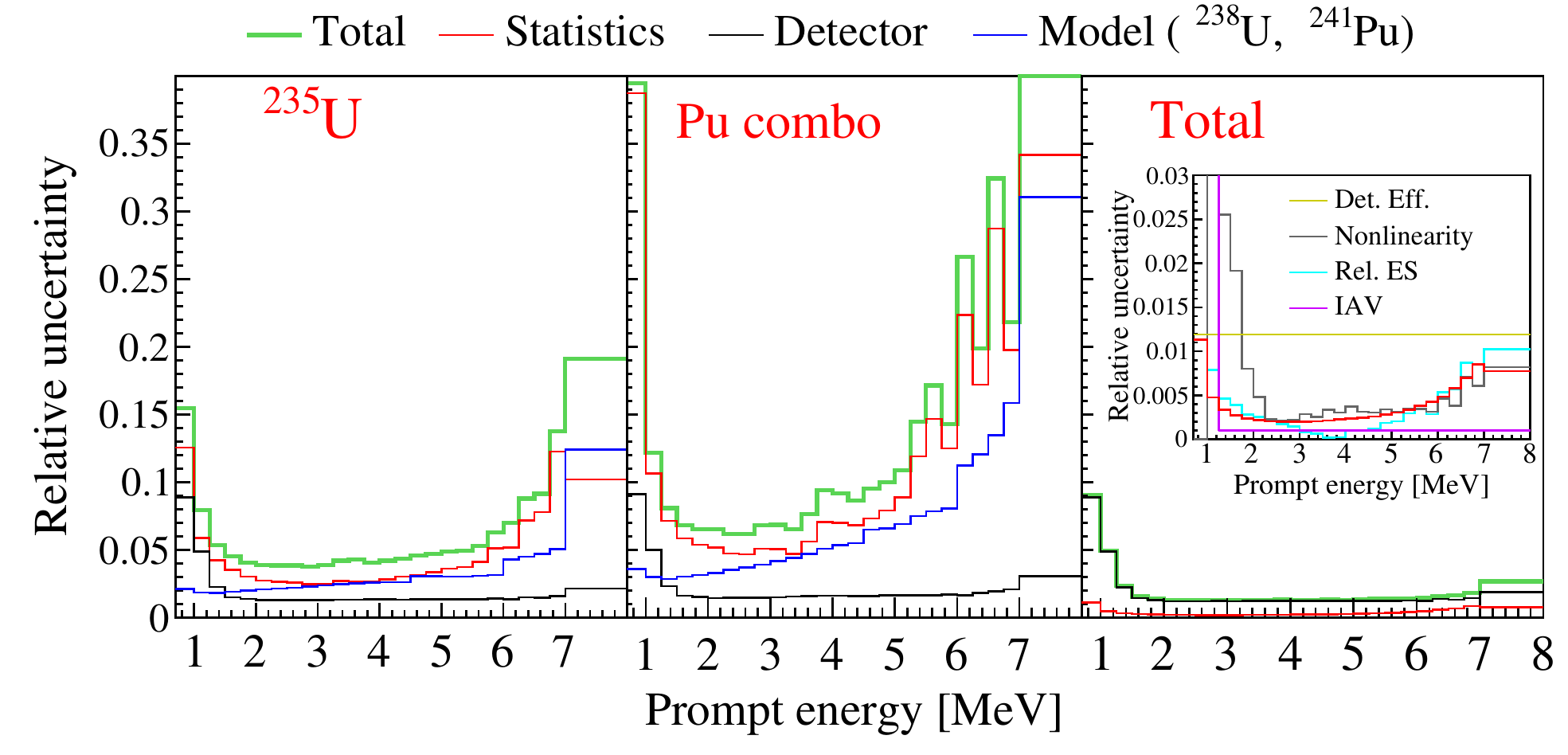}
\figcaption{\label{Fig-PromptErrBudget}  Dominant components of the energy-dependent uncertainties for isotopic (\UL~and Pu combo) and total prompt energy spectra. 
Here relative uncertainty represents the square root of the diagonal elements of the covariance matrix. 
``Detector'' uncertainty (detector response uncertainty) contains the contribution of the uncertainties from detection efficiency (``Det. Eff.''), the energy nonlinearity model (``Nonlinearity''), energy scale difference between ADs (``Rel. ES''), and the IAV effect (``IAV''). ``Model (\UH, \PuH)'' uncertainty represents uncertainties of the models for \UH~and \PuH~in the analysis to extract the isotopic (\UL~and Pu combo) prompt energy spectra. The inset shows the components of the detector response uncertainty for the total prompt energy spectrum, with statistical uncertainty shown for comparison. Other uncertainties (e.g. background uncertainty) have a small contribution to the total uncertainty and are not shown. }
\end{center}

The correlation matrix of the isotopic and total prompt energy spectra is shown in Fig.~\ref{Fig-TotalCorrelation}. 
The total uncertainty including both the rate and spectral shape of the total prompt energy spectrum is $\sim$1.3\% in 2 to 5~MeV energy region, with large bin-to-bin correlation due to the dominant detection efficiency uncertainty. 
The relative uncertainty in the spectral shape is smaller ($<$0.5\% in 2-5~MeV energy region) than the overall uncertainty for the total prompt energy spectrum~\cite{Adey:2019ywk}.
The correlation between the same bin of the prompt energy spectra for \UL~and Pu combo is mostly less than 0 since the isotopic prompt energy spectra constitute the total measurement and their statistical fluctuations are anticorrelated. 

\begin{center}
\includegraphics[width=0.5\columnwidth]{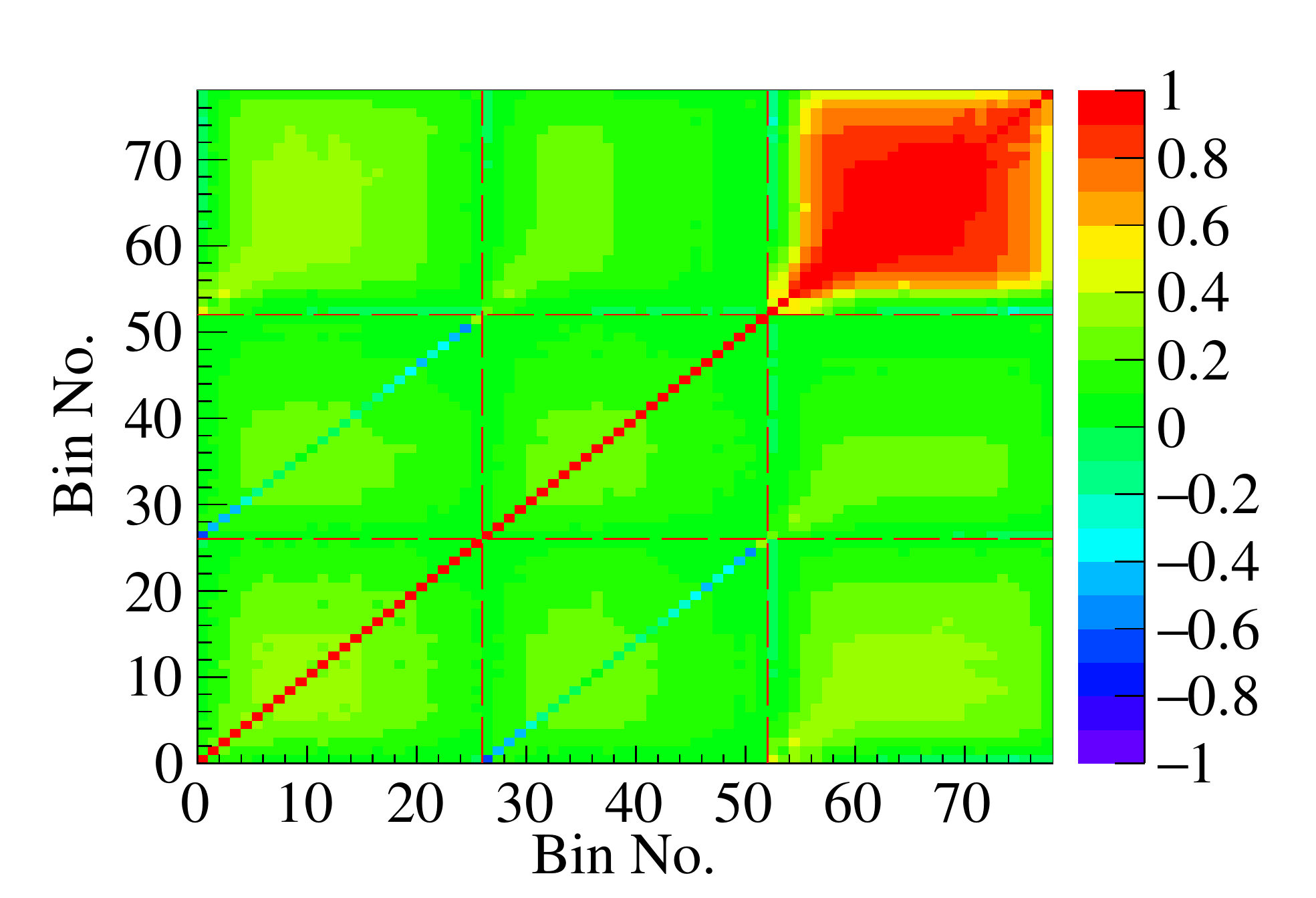}
\figcaption{\label{Fig-TotalCorrelation}  Correlation matrix of the isotopic (\UL~and Pu combo) and total prompt energy spectra. The first, middle, and last 26 bins of the correlation matrix represent the prompt energy spectra of ${}^{235}$U, Pu combo, and total measurement in Fig.~\ref{Fig-PromptErrBudget} respectively. }
\end{center}

The total prompt energy spectrum and the individually extracted isotopic prompt energy spectra of the IBD reaction can be compared with antineutrino models after taking into account the detector response, and are used to obtain the generic antineutrino energy spectra of the IBD reaction in Sec.~{\ref{Sec-Unfolding}}.

The systematic uncertainty dominates the total uncertainty for the total prompt energy spectrum with the bin width of 0.25~MeV as shown in Fig.~\ref{Fig-PromptErrBudget}. Since the statistical uncertainties are within sub-percent level at all energies, the bin width is further reduced to better match the energy resolution of the Daya Bay detectors. This can also provide more spectral information, thus more studies can be carried out. 
The Appendix contains a discussion of the measurement and applications of the finely-binned prompt energy spectrum. 

\section{Generic antineutrino energy spectra of the IBD reaction}
\label{Sec-Unfolding}
To provide a model-independent reactor \nuebar~energy spectrum prediction, unfolding techniques are used to obtain generic antineutrino energy spectra of the IBD reaction (or generic \nuebar~energy spectrum weighted by the IBD cross section). Unfolding transforms the prompt energy spectrum to the \nuebar~energy spectrum for direct comparison with spectra from other experiments as well as with theoretical predictions. Daya Bay previously used the SVD regularization method to produce the first generic \nuebar~energy spectrum~\cite{An:2016srz}. In this Section, unfolding using a new method, the Wiener-SVD method~\cite{Tang:2017rob}, is described and used to produce total and isotopic (\UL~and Pu combo) \nuebar~energy spectra. The spectra are compared to results obtained with the SVD~\cite{Hocker:1995kb} and Bayesian iteration~\cite{DAgostini:1994fjx} methods.

Because of the finite energy resolution, the response matrix is found to be ill-conditioned with close-to-zero singular values. 
If the inverse of the detector response matrix is naively used to obtain the \nuebar~energy spectrum, statistical fluctuations and systematic variations will be significantly amplified. This phenomenon would lead to meaningless results. 
Thus the unfolding techniques are utilized to suppress the fluctuations and solve this problem. 
In this work, the unfolded results of Bayesian iteration method and SVD method are based on an unfolding package RooUnfold~\cite{Adye:2011gm}, while the results of Wiener-SVD method are based on the algorithm shared in the GitHub~\cite{Tang:2017rob}.  
The Bayesian iteration method needs an initial guess of the \nuebar~energy spectrum as a starting value and the spectrum is updated iteratively based on the Bayes' theorem, which takes into account the response matrix and the observed prompt energy spectrum. The iteration is stopped when the change of the unfolded spectrum is small and before the unphysical fluctuations appear~\cite{DAgostini:1994fjx}. 
The SVD method is based on matrix inversion calculation, reducing or suppressing the small singular values in the detector response matrix with a regularization parameter. This method is improved with $a~priori$ information about the solution to overcome the instability of unfolding~\cite{Hocker:1995kb}. 
The Wiener-SVD approach achieves the unfolding by maximizing the signal to noise ratio in the effective frequency domain (a Wiener filter constructed based on an input \nuebar~model and the observed spectrum) thus avoiding the need for a regularization parameter used by other methods. It also has smaller mean square error (MSE, which averages the total bias and variance) than traditional SVD regularization method~\cite{Tang:2017rob}. 

As described above, a model of \nuebar~energy spectrum is needed as an input in each of these unfolding procedures. As shown in Refs. \cite{Ashenfelter:2018jrx,Adey:2019ywk,Andriamirado:2020erz,AlmazanMolina:2020jlh}, both \UL~and \PuL~are likely to be responsible for the spectral distortion, rather than only the \UL~contribution. 
In this case, the Huber-Mueller model with spectral correction is used as an input model. 
The correction of the spectral distortion is the ratio of the total measurement of the prompt energy spectrum over the prediction based on the Huber-Mueller model at Daya Bay. 
The ratio defined above is based on prompt energy, and the prompt energy ($E_{\rm p}$) can be shifted to \nuebar~energy ($E_{\bar{\nu}_e}$) using the following formula: $E_{\bar{\nu}_e}\approx E_{\rm p}+0.78~\text{MeV}$. 
The ratio obtained above is applied to all Huber-Mueller model predictions (total, \UL, and Pu combo \nuebar~energy spectra) to construct an input model for the unfolding method. 
However, the input model is not identical to the true \nuebar~spectrum, and this effect will induce additional bias in the unfolded results. 
Unfolding methods suppress the fluctuations as well as the fine structures in the original \nuebar~energy spectra. 
With different variations added on the input model, changes in the unfolded results are observed. 
A study was done to consider the effect of fine structures based on the SM2018 model~\cite{Estienne:2019ujo} which found a $\sim$2\% variation compared with a smooth spectrum (smearing with the energy resolution of Daya Bay detectors). This variation is consistent with the sawtooth distortions found in Ref.~\cite{Danielson:2018tzi}. 
Based on this study, the possible variation due to the difference between the input model and the true spectrum is set to 3\% (bin-to-bin uncorrelated) conservatively. 

The relative performance of these methods also depends on the specific convergence criteria applied.
The Bayesian iteration method needs the number of iterations as an input, while the SVD method needs input for the regularization parameter. To optimize the input parameters for these methods and compare the MSE between these methods, dedicated tests with toy Monte Carlo simulations are done. 
In the tests, \nuebar~energy spectra based on the Huber-Mueller model ($\boldsymbol{S}^{\bar{\nu}_e}$) are used to generate 10000 samples of total and isotopic prompt energy spectra with the response matrix and fluctuations based on the covariance matrix in Sec.~{\ref{Sec-CorrelationAna}}. 
The total and isotopic unfolded \nuebar~energy spectra ($\boldsymbol{S}^{\rm unfold}$) are obtained with each unfolding method for each sample.
Based on a dedicated study on the response matrix, the binning method for the unfolded \nuebar~energy spectra is optimized with totally 25 energy bins, while originally the prompt energy spectra have 26 energy bins in total.
For each toy Monte Carlo test, total and isotopic \nuebar~energy spectra are unfolded individually, but they are combined to one spectrum in order to evaluate their correlation more conveniently. 
Then $\boldsymbol{S}^{\rm unfold}$ are compared to the true \nuebar~energy spectra ($\boldsymbol{S}^{\bar{\nu}_e}$) to evaluate the bias and uncertainty of the unfolded result. 
Here $\boldsymbol{S}^{\rm unfold}$ and $\boldsymbol{S}^{\bar{\nu}_e}$ are defined to contain the total, \UL~and Pu combo spectra, with 75 (=25$\times$3) energy bins chaining all these three spectra together (each with 25 energy bins).
In each unfolding procedure, a 3\% bin-to-bin uncorrelated fluctuation is added on the true \nuebar~energy spectra to construct different input \nuebar~spectra, in order to account for the possible variation from input models. 
For input \nuebar~isotopic spectra, an additional 10\% systematic fluctuation (bin-to-bin correlated) is added to account for the possible size of the spectral distortion. 
An element of the covariance matrix is calculated by
\begin{align}
\boldsymbol{\rm Cov}_{ij}=\frac{1}{N}\sum_t^N\frac{(\boldsymbol{S}^{\rm unfold}_{ti}-\boldsymbol{S}^{\bar{\nu}_e}_{i})(\boldsymbol{S}^{\rm unfold}_{tj}-\boldsymbol{S}^{\bar{\nu}_e}_{j})}{\boldsymbol{S}^{\bar{\nu}_e}_{i}\boldsymbol{S}^{\bar{\nu}_e}_{j}},
\label{Eq-CovBias}
\end{align}
where $t$ is the sample number and runs from 1 to $N$ = 10000, $i$ represents the $i$-th energy bin in the spectra. 
$\boldsymbol{\rm Cov}$ contains the effect of the bias from the input model and the uncertainty from the measurement. 
The best number of iterations for the Bayesian iteration method is determined to be 1 while the regularization parameter for the SVD method is determined to be 21, with the least MSE for toy Monte Carlo tests. Here the MSE is defined as the summation of the square root of the diagonal terms in $\boldsymbol{\rm Cov}$. 
$\boldsymbol{\rm Cov}$ contains the information of the correlation among total and isotopic unfolded \nuebar~energy spectra. 

Figure~\ref{Fig-UnfoldingResults} shows the unfolded \nuebar~energy spectra as well as their relative uncertainties. 
The total uncertainties of the unfolded \nuebar~energy spectra from the Wiener-SVD method in the 3 to 6~MeV energy region are 2\% for the total \nuebar~energy spectrum and 3.5\% (5\%) for the \UL~ (Pu combo) \nuebar~energy spectrum. 
The Bayesian iteration and SVD methods yield consistent results.  
The result from the Wiener-SVD method has the least MSE in the 3 to 6~MeV energy region, especially for the total \nuebar~energy spectrum. 
This is presumably due to the optimized signal to noise ratio in the effective frequency domain for the Wiener filter, as found in Ref.~\cite{Tang:2017rob}. 
Since the Wiener filter has small suppression in the low effective frequency domain, the result from the Wiener-SVD method has larger fluctuation and bigger uncertainties than traditional Bayesian iteration method and SVD method in the low and high energy regions, where the original statistical fluctuations are large. 

\begin{center}
\includegraphics[width=\columnwidth]{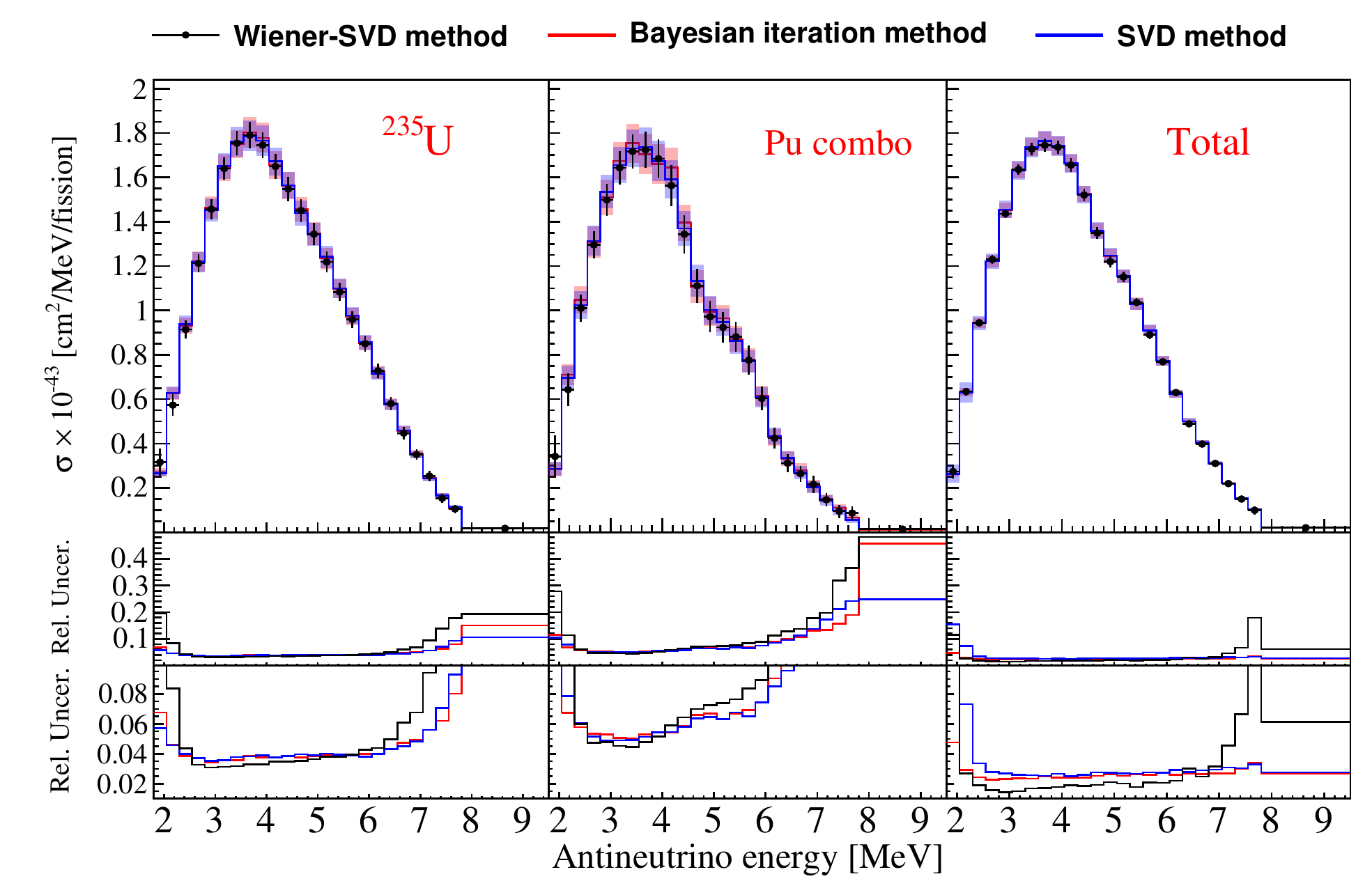}
\figcaption{\label{Fig-UnfoldingResults}  (Top panels) Unfolded isotopic (\UL~and Pu combo) and total \nuebar~energy spectra weighted by IBD cross section. The error bars on the data points represent the square root of diagonal elements of the covariance matrix, which is generated by Monte Carlo methods. (Middle panels) Relative uncertainty of the unfolded isotopic and total \nuebar~energy spectra. (Bottom panels) Enlarged plot of the relative uncertainty of the unfolded isotopic and total \nuebar~energy spectra. The result from the Wiener-SVD method has the smallest MSE in the 3 to 6~MeV energy region, especially for the total \nuebar~energy spectrum. }
\end{center}

To gain insight on the contribution from each component of the energy-dependent uncertainties to the unfolded \nuebar~energy spectra, dedicated tests with toy Monte Carlo simulations were done. 
Toy Monte Carlo samples of total and isotopic prompt energy spectra were generated with the response matrix and fluctuations based on the error budget in Fig.~\ref{Fig-PromptErrBudget}. 
In each unfolding procedure, the input \nuebar~energy spectra are exactly the same as the true \nuebar~energy spectrum for all toy Monte Carlo tests. 
The total and isotopic unfolded \nuebar~energy spectra ($\boldsymbol{S}^{\rm unfold}$) are obtained with each unfolding method and are used to calculate elements of the covariance matrix based on the following formula: 
\begin{align}
\boldsymbol{\rm Cov}_{ij}^{\prime}=\frac{1}{N}\sum_t^N\frac{(\boldsymbol{S}^{\rm unfold}_{ti}-\bar{\boldsymbol{S}}^{\rm unfold}_{i})(\boldsymbol{S}^{\rm unfold}_{tj}-\bar{\boldsymbol{S}}^{\rm unfold}_{j})}{\bar{\boldsymbol{S}}^{\rm unfold}_{i}\bar{\boldsymbol{S}}^{\rm unfold}_{j}}. 
\label{Eq-Cov}
\end{align}
Here $\bar{\boldsymbol{S}}^{\rm unfold}=\frac{1}{N}\sum_t^N{\boldsymbol{S}^{\rm unfold}_{t}}$ is the average of the unfolded \nuebar~energy spectra $\boldsymbol{S}^{\rm unfold}$. 
With toy Monte Carlo samples incorporating fluctuations based on the covariance matrixes of different components in Fig.~\ref{Fig-PromptErrBudget}, contributions from the uncertainties due to statistics, detector, and model (\UH, \PuH) to the total uncertainties of the unfolded results are obtained. 
The total covariance matrix without the bias from the input model ($(\boldsymbol{\rm Cov}^{\prime})^{\rm total}$) is obtained based on toy Monte Carlo samples taking into account the total uncertainty from the measurement, using Eq.~\ref{Eq-Cov}. 
Since the covariance matrix ($\boldsymbol{\rm Cov}^{\rm total}$) from Eq.~\ref{Eq-CovBias} takes into account the bias from the input model, the additional bias from the input model ($\boldsymbol{\rm Cov}^{\rm bias}$) is calculated by the following equation: 
\begin{align}
\boldsymbol{\rm Cov}^{\rm bias}=\boldsymbol{\rm Cov}^{\rm total} - (\boldsymbol{\rm Cov}^{\prime})^{\rm total}. 
\end{align}
Based on these tests, the dominant components of the energy-dependent uncertainties for total and isotopic unfolded \nuebar~energy spectra based on the Wiener-SVD method are shown in Fig.~\ref{Fig-NeutrinoErrBudget}. 
Contributions from statistics and the models for \UH~and \PuH~dominate the total uncertainties for isotopic spectra. 
The input model used by the unfolding method and the detector response uncertainty both induce an uncertainty on the total unfolded \nuebar~energy spectrum at the 1\% level and dominate the total uncertainty in the spectrum.

The unfolded \nuebar~energy spectra are the \nuebar~energy distributions weighted by IBD cross section for one fission reaction and the subsequent beta decay reactions. 
They can thus be directly compared with theoretically predicted \nuebar~energy spectra of the IBD reaction. 
The unfolded $\bar\nu_e$ energy spectra, both with and without IBD cross section weighting, are provided in the supplemental material.

\begin{center}
\includegraphics[width=\columnwidth]{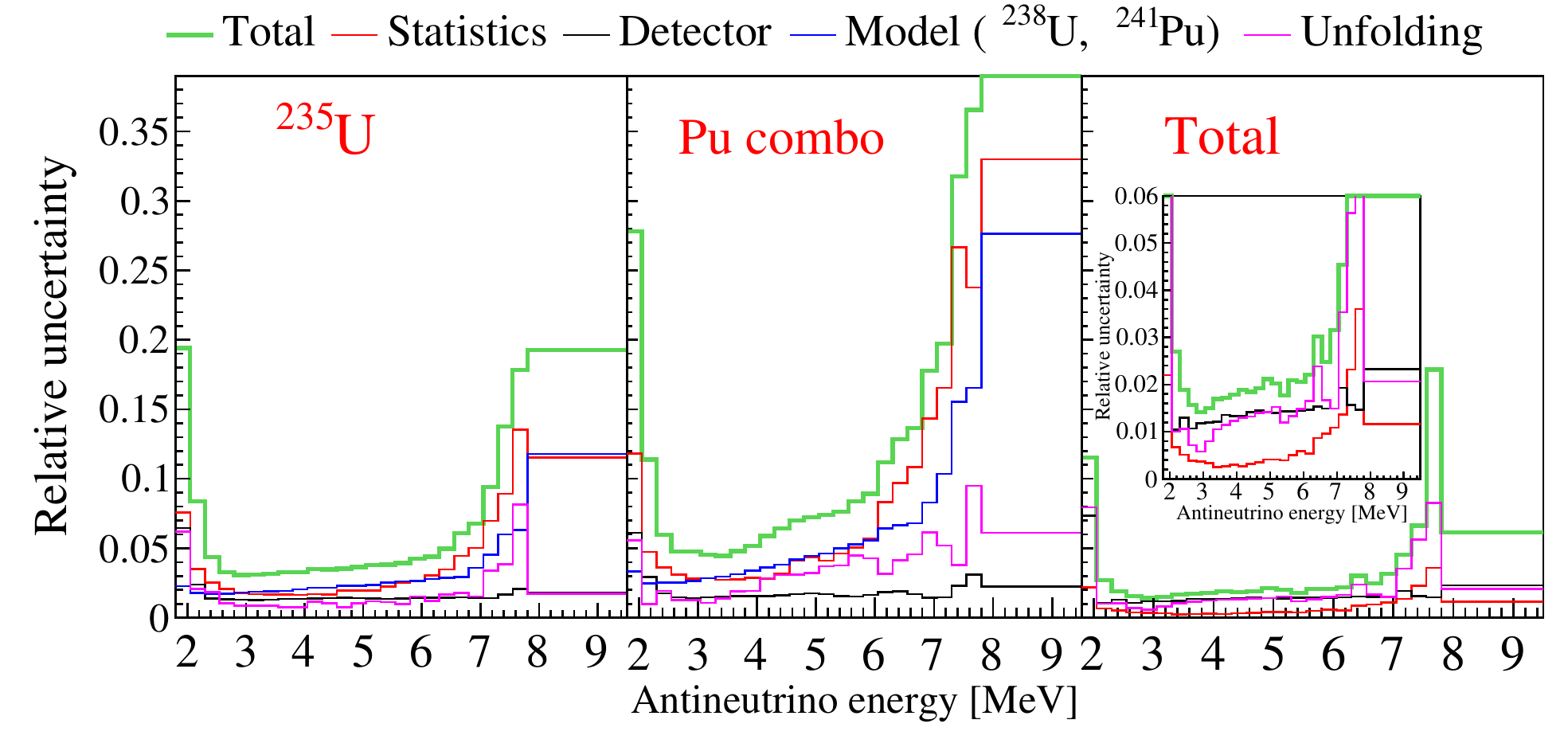}
\figcaption{\label{Fig-NeutrinoErrBudget}  Dominant components of the energy-dependent uncertainties for isotopic (\UL~and Pu combo) and total unfolded \nuebar~energy spectra. 
The inset shows the details of the energy-dependent uncertainties for the total unfolded \nuebar~energy spectrum. 
``Detector'' uncertainty (detector response uncertainty) represents the contribution of the uncertainties from detection efficiency, energy scale difference between ADs, energy nonlinearity model, and IAV effect. Model (\UH, \PuH) uncertainty represents uncertainties of the input model in the analysis to extract the isotopic (\UL~and Pu combo) prompt energy spectra. The contribution of the uncertainty from unfolding method takes into account the bias from the fluctuations of the input model. }
\end{center}

\section{Application of the generic antineutrino energy spectra}
\label{Sec-Application}
As introduced above, the unfolded \nuebar~energy spectra of the IBD reaction can provide a model-independent input for other reactor antineutrino experiments. Furthermore, it can serve as an additional resource for validating theoretical models and standard nuclear databases. 
In this section, the application of the unfolded \nuebar~energy spectra is shown with detailed examples. 

\subsection{Prediction for other reactor antineutrino experiments}
The generic unfolded \nuebar~energy spectra of the IBD reaction can be directly used by other reactor antineutrino experiments utilizing the IBD reaction for \nuebar~detection. 
Experiments which do not utilize the IBD reaction can in principle remove the IBD cross section to obtain original \nuebar~spectra from beta decay of fission fragments in the reactor. 
The antineutrino energy spectrum of the IBD reaction for an experiment with fission fractions ($f_{235}^A$, $f_{238}^A$, $f_{239}^A$, $f_{241}^A$) can be predicted as: 
\begin{align} 
\label{Formula-Prediction}
\boldsymbol{S}_A&=\boldsymbol{S}_{\rm total}+\Delta f_{235}\boldsymbol{S}_{235}+\Delta f_{239}\boldsymbol{S}_{239}+\Delta f_{238}\boldsymbol{S}_{238}+\Delta f_{241}\boldsymbol{S}_{241}\nonumber\\
&=\boldsymbol{S}_{\rm total}+\Delta f_{235}\boldsymbol{S}_{235}+\Delta f_{239}\boldsymbol{S}_{\rm combo}+\Delta f_{238}\boldsymbol{S}_{238}+(\Delta f_{241}-0.183\times \Delta f_{239})\boldsymbol{S}_{241}.  
\end{align}
Here $A$ represents an arbitrary reactor antineutrino experiment; $\Delta f_i$ is the fission fraction difference between experiments for $i$-th fissile isotope, $\Delta f_i= f^{A}_i-f^{DB}_i$, where $f^{DB}$ is the average effective fission fraction of the Daya Bay experiment (Section~\ref{Sec-CorrelationAna}); $\boldsymbol{S}_{\rm total}$, $\boldsymbol{S}_{235}$, and $\boldsymbol{S}_{\rm combo}$ are the total and isotopic unfolded \nuebar~energy spectra respectively. When the other experiment has different fission fractions than Daya Bay, small corrections based on the theoretical model inputs of $\boldsymbol{S}_{238}$ and $\boldsymbol{S}_{241}$ are necessary. 
For experiments detecting the \nuebar~from commercial LEU reactors, the Pu combo spectrum is recommended since it has smaller uncertainty and the fission fractions of \PuL~and \PuH~are strongly correlated. 

To estimate the uncertainty of the prediction using Eq.~\ref{Formula-Prediction}, one should consider the correlation between total and isotopic spectra. 
Equation~\ref{Formula-Prediction} is converted to a matrix version in order to calculate the covariance matrix of the prediction more easily. A transformation matrix $\boldsymbol R$ is constructed as follows:

\begin{align} 
{\boldsymbol R}&=
\left(\begin{array}{c|c|c|c|c} {\boldsymbol I}_{25} & \Delta f_{235}{\boldsymbol I}_{25} & 
\Delta f_{239}{\boldsymbol I}_{25} & 
\Delta f_{238}{\boldsymbol I}_{25} & 
 \left(\Delta f_{241}-0.183 \times \Delta f_{239}\right){\boldsymbol I}_{25} \end{array} \right) \nonumber\\
&=\left(\begin{array}{ccc|ccc|ccc|ccc|ccc}
1 & & & \Delta f_{235} & & & \Delta f_{239}& & & \Delta f_{238} & & & \cdots & & \\
 & \ddots & & & \ddots & & & \ddots & & & \ddots & & & \ddots &\\ 
 & & 1 & & & \Delta f_{235} & & & \Delta f_{239} & & & \Delta f_{238} & & & \cdots \\
\end{array}\right).
\end{align} 

Here $\boldsymbol R$ is a $25\times 125$ matrix, transforming 5 spectra into a prediction with different fission fractions; $\boldsymbol{I}_{a}$ means the $a\times a$ identity matrix. 
The predicted \nuebar~energy spectrum corrected for fission fractions can be obtained by
\begin{align} 
\boldsymbol{S}_{\rm pred}={\boldsymbol{R}}\cdot\begin{pmatrix} {\boldsymbol{S}}_{\rm total} \\ {\boldsymbol{S}}_{235} \\{\boldsymbol{S}}_{\rm combo} \\ {\boldsymbol{S}}_{238} \\ {\boldsymbol{S}}_{241} \end{pmatrix}. 
\end{align} 
In the latter example, ${\boldsymbol S}_{238}$ is chosen to be the Mueller \UH~model~\cite{Mueller:2011nm} with a 15\% bin-to-bin uncorrelated uncertainty and ${\boldsymbol S}_{241}$ is chosen to be the Huber \PuH~model~\cite{Huber:2011wv} with a 10\% bin-to-bin uncorrelated uncertainty. 
They both have little contribution to the total predicted spectrum because the fission fraction of \UH~is relatively stable over time and the ratio between the fission fractions of \PuL~and \PuH~is normally close to 0.183. Other models for \UH~and \PuH~with reasonable uncertainties could be chosen as well. 
The uncertainties of the prediction can be obtained by error propagation. The covariance matrix ${\boldsymbol {\rm Cov}}_{\rm pred}$ of $\boldsymbol S_{\rm pred}$ is calculated by the formula:
\begin{align} 
{\boldsymbol {\rm Cov}}_{\rm pred}={\boldsymbol R}\cdot{\boldsymbol {\rm Cov}}_{\rm unfold}\cdot{\boldsymbol R}^T.
\end{align} 
Here ${\boldsymbol {\rm Cov}}_{\rm unfold}$ is constructed based on the covariance matrix of the unfolded \nuebar~energy spectra from Eq.~\ref{Eq-CovBias}, with the correlation between total and isotopic spectra taken into account. In addition, ${\boldsymbol {\rm Cov}}_{\rm unfold}$ also includes the covariance matrix of the models for \UH~and \PuH, without consideration on their correlation with other spectra. 
In this way, the prediction of the \nuebar~energy spectrum for the other experiment can be obtained. 
Then the \nuebar~energy spectrum can be converted to the measurement of the prompt energy spectrum with detector response of the other experiment taken into consideration. 

To demonstrate how to utilize the unfolded \nuebar~energy spectra and calculate the uncertainty of the prediction, examples are shown for predictions with different fission fractions than Daya Bay. 
The measurement of the prompt energy spectrum at EH1 AD1 is divided into 20 groups ordered by the \PuL~effective fission fraction in each week, as was done in Ref.~\cite{Adey:2019ywk}. 
The data in different groups represent different periods corresponding to different reactor burnup.  
Predictions of the prompt energy spectrum at EH1 AD1 in the earlier period ($5^{\rm th}$ group, $f_{235}:f_{239}:f_{238}:f_{241}=0.597: 0.278: 0.076: 0.049$), intermediate period ($10^{\rm th}$ group, $f_{235}:f_{239}:f_{238}:f_{241}=0.567: 0.301: 0.076: 0.056$) and later period ($15^{\rm th}$ group, $f_{235}:f_{239}:f_{238}:f_{241}=0.541: 0.322: 0.076: 0.061$) are performed. 
The predictions of the measured prompt energy spectrum based on unfolded \nuebar~energy spectra are obtained by applying the response matrix to transform the \nuebar~energy to the prompt energy of IBD events. 
Since the prompt energy spectra for Daya Bay are predicted, we can also do a prediction based on total and extracted isotopic prompt energy spectra $\boldsymbol{S}_{\rm prompt}$ in Sec.{~\ref{Sec-CorrelationAna}} directly for comparison. The results based on $\boldsymbol{S}_{\rm prompt}$ can be used to validate the prediction from unfolded \nuebar~energy spectra. 
The detector information (detector DAQ live time, detection efficiency,  etc.), reactor information (time-varying thermal power, etc.), and survival probability from neutrino oscillation are taken into account in the prediction. 
The non-equilibrium effect is also considered in the prediction based on the reactor running status as was done in Ref.~\cite{Adey:2019ywk}. 
The predictions based on $\boldsymbol{S}_{\rm unfold}$ and $\boldsymbol{S}_{\rm prompt}$ are shown in Fig.~\ref{Fig-PredictionExamples}. 
The measurements of the prompt energy spectrum with statistical uncertainty at EH1 AD1 are also shown in Fig.~\ref{Fig-PredictionExamples} for comparison with $\sim$40,000 IBD events excluding backgrounds in each period. 
The difference between the predictions based on $\boldsymbol{S}_{\rm unfold}$ and $\boldsymbol{S}_{\rm prompt}$ is small and well within uncertainties. The predictions are also consistent with the measurement in the 1 to 7~MeV energy region within statistical uncertainties. 
At lower or higher energy region the predictions are also consistent with the measurement when considering systematic uncertainties. 
This suggests the feasibility of utilizing unfolded \nuebar~energy spectra to get prediction for other reactor antineutrino experiments. 
The application to the experiments exposed to a larger range of fission fractions can be done in a similar way. 

\begin{center}
\includegraphics[width=\columnwidth]{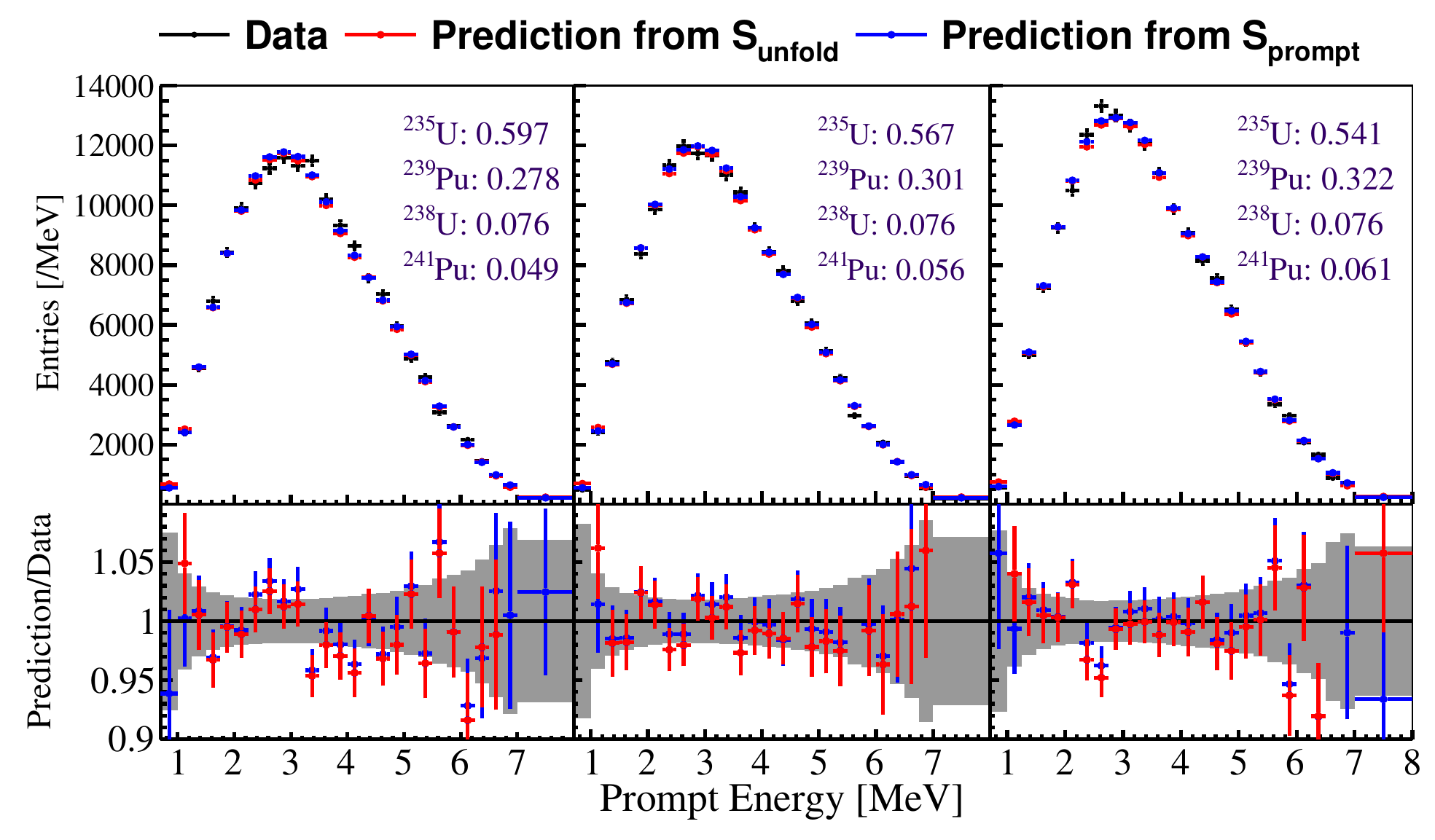}
\figcaption{\label{Fig-PredictionExamples}  (Top) Prediction of the prompt energy spectra with different fission fractions at Daya Bay based on unfolded \nuebar~energy spectra $\boldsymbol{S}_{\rm unfold}$ or prompt energy spectra $\boldsymbol{S}_{\rm prompt}$. The measurements with corresponding fission fractions from EH1 AD1 are shown for comparison with $\sim$40,000 IBD events in each period. Only statistical uncertainty on the measurement is shown in the plot. The legends on the top right represent the effective fission fractions of \UL, \PuL, \UH, \PuH~for each period respectively. (Bottom) Comparison between prediction and measurement at Daya Bay. The error bars on the histogram only contain the uncertainties from prediction. Statistical uncertainty for the measurement is shown with the grey band. The difference between the prediction and the data are consistent within statistical uncertainty in the 1 to 7~MeV energy region. }
\end{center}


\subsection{Comparison with theoretical reactor antineutrino models}
Another application of the unfolded \nuebar~spectra is the comparison with theoretical reactor antineutrino models directly. 
Since the Huber-Mueller model provides the uncertainty and has a long history of comparison with measurements, this model is used as an example to compare with the unfolded \nuebar~energy spectra from Sec.~{\ref{Sec-Unfolding}}. 
The detailed comparison between the Huber-Mueller model and the latest measurement of the prompt energy spectrum is shown in Ref.~\cite{Adey:2019ywk}. 
Here the focus is on the comparison based on \nuebar~energy spectra. 
The covariance matrix of the Huber-Mueller model ($\boldsymbol{\rm Cov}^{\rm HM}$) is obtained by doing toy Monte Carlo based on the bin-to-bin correlated uncertainty and uncorrelated uncertainties provided in Refs.~\cite{Mueller:2011nm,Huber:2011wv}, and the calculation method is the same as in previous publications~\cite{An:2016srz,Adey:2019ywk}. 
The difference between the Huber-Mueller model ($\boldsymbol{S}^{\rm HM}$) and unfolded \nuebar~energy spectra ($\boldsymbol{S}^{\rm unfold}$) is evaluated based on the $\chi^2$ value defined as:
\begin{align}
    \chi^2=\sum_{i,j}(\boldsymbol{S}_i^{\rm unfold}-\boldsymbol{S}_i^{\rm HM})(\boldsymbol{\rm Cov}^{\rm all})^{-1}_{ij}(\boldsymbol{S}_j^{\rm unfold}-\boldsymbol{S}_j^{\rm HM}),
\end{align}
where $\boldsymbol{S}^{\rm unfold}$ represents the \UL, Pu combo, or total unfolded \nuebar~energy spectrum (each with 25 bins in this case), and $\boldsymbol{\rm Cov}^{\rm unfold}$ represents their respective covariance matrix; $\boldsymbol{S}^{\rm HM}$ represents the prediction of the total or isotopic \nuebar~energy spectrum based on the Huber-Mueller model (each with 25 bins). $\boldsymbol{\rm Cov}^{\rm all}$ is the total covariance matrix of $\boldsymbol{S}^{\rm unfold}$ and $\boldsymbol{S}^{\rm HM}$, which is the sum of $\boldsymbol{\rm Cov}^{\rm unfold}$ and $\boldsymbol{\rm Cov}^{\rm HM}$. The $\chi^2$ is equal to 57.2, 12.3, and 98.5 for \UL, Pu combo, and total \nuebar~energy spectra respectively. Since both the rate and spectral shape of the \nuebar~energy spectra are compared, the number of degrees of freedom in the above comparison is exactly the number of bins (25). Thus, the corresponding $p$-values are 2.52$\times 10^{-4}$, 0.984, and 1.14$\times 10^{-10}$. The corresponding significance of deviation for the total \nuebar~energy spectrum is 6.4$\sigma$, confirming the flux and spectral difference between measurement and the Huber-Mueller model. 

Since the bias from input models in the unfolding method is included when evaluating the covariance matrix of the unfolded \nuebar~energy spectra, 
the sensitivity of the unfolded \nuebar~energy spectra is less powerful for reactor \nuebar~spectrum model discrimination than the original measurements of prompt energy spectrum.
To achieve a more discriminating comparison with the Daya Bay data by avoiding the impact of unfolding bias, another method can be used to do the comparison. 
In principle, the unfolded \nuebar~energy spectrum with the Wiener-SVD unfolding method ($\boldsymbol{S}^{\rm Wiener}$) is obtained based on the following formula:
\begin{align}
\boldsymbol{S}^{\rm Wiener}=\boldsymbol{A}_{c}\boldsymbol{M}^{-1}\boldsymbol{S}^{\rm prompt}.  
\end{align}
Here $\boldsymbol{M}$ is the response matrix of the detector; $\boldsymbol{S}^{\rm prompt}$ is the measurement from the detector; $\boldsymbol{A}_c$ is a smearing matrix used during the Wiener-SVD unfolding procedure, and it is constructed based on the signal to noise ratios in the effective frequency domain given expectations of signal and noise.
The smearing matrix $\boldsymbol{A}_{c}$ suppresses the unphysical fluctuations but introduces biases as well during the unfolding procedure.
The covariance matrix of the unfolded \nuebar~energy spectrum ($\boldsymbol{\rm Cov}^{\rm Wiener}$) is calculated by error propagation from the Wiener-SVD unfolding package. 
$\boldsymbol{\rm Cov}^{\rm Wiener}$ does not include the bias from the unfolding method, and it is different from $\boldsymbol{\rm Cov}^{\rm unfold}$ based on toy Monte Carlo tests in Sec.~{\ref{Sec-Unfolding}}. 
A more detailed description is found in Ref.~\cite{Tang:2017rob}. 
The smearing matrices $\boldsymbol{A}_c$ are provided in the supplemental material.

To compare with the unfolded \nuebar~energy spectra, the theoretical \nuebar~energy spectra are smeared in the same way as the Wiener-SVD unfolding method: $\boldsymbol{S}^{\rm smear}=\boldsymbol{A}_c\boldsymbol{S}^{\rm HM}$. The covariance matrix of $\boldsymbol{S}^{\rm smear}$ is calculated by $\boldsymbol{\rm Cov}^{\rm smear}=\boldsymbol{A}_c\boldsymbol{\rm Cov}^{\rm HM}\boldsymbol{A}_c^T$. 
In this way, the $\chi^2$ between the Huber-Mueller model and the unfolded \nuebar~energy spectra is calculated using the following formula:
\begin{align}
\label{Eq-UnfoldCompare}
    \chi^2=\sum_{i,j}(\boldsymbol{S}_i^{\rm unfold}-\boldsymbol{S}_i^{\rm smear})(\boldsymbol{\rm Cov}^{\rm sum})^{-1}_{ij}(\boldsymbol{S}_j^{\rm unfold}-\boldsymbol{S}_j^{\rm smear}),
\end{align}
where $\boldsymbol{\rm Cov}^{\rm sum}=\boldsymbol{\rm Cov}^{\rm Wiener}+\boldsymbol{\rm Cov}^{\rm smear}$. Based on the above procedure, the $\chi^2$ is equal to 64.9, 22.3, and 108.8 for \UL, Pu combo, and total \nuebar~energy spectra, respectively. The corresponding $p$-values are 2.11$\times 10^{-5}$, 0.618, 1.99$\times 10^{-12}$, indicating greater discrepancy between the measurements and the model than considering the impact of bias in the unfolded \nuebar~energy spectra. 
This conclusion is further confirmed by a statistical check based on a dedicated toy Monte Carlo test. 

The inclusion of the smearing matrix in the comparison with theoretical models maintains the discriminating sensitivity of the original data by avoiding the impact of bias in the unfolded \nuebar~energy spectra. 
This method can also be used to discriminate different hypotheses, with the same sensitivity of discrimination based on the original measurements of prompt energy spectra. 
To validate the feasibility of this method, two different hypotheses are considered: the ``normalized Huber-Mueller model'' with the same flux as the original measurements, and the ``normalized distorted Huber-Mueller model'' with an additional Gaussian (maximum: 10\%, mean: 6~MeV, sigma: 0.6~MeV) added on the Huber-Mueller model, with the flux normalized as well. 
Different hypotheses are compared with the unfolded \nuebar~energy spectra based on the Eq.~\ref{Eq-UnfoldCompare} (method 1), while the comparison between data and models based on the prompt energy spectra (method 2) is also done to verify the method 1. 
The $\chi^2$ based on \UL~spectrum with method 1 (method 2) for the ``normalized Huber-Mueller model'' is 45.0 (45.1), while for the ``normalized distorted Huber-Mueller model'' the $\chi^2$ is 19.3 (19.3). 
To quantify the comparison between these two hypotheses, a frequentist approach is used based on 10000 toy experiments generated for the ``normalized distorted Huber-Mueller model'' hypothesis. 
A probability distribution function (PDF) is formed from the $\Delta \chi^2$ values for each toy experiment between the two models for both methods. 
The observed experimental $\Delta \chi^2$ value is compared to its respective PDF to determine quantitative preference of the observation for a particular model. 
The $\Delta \chi^2$ is found to be 25.7 (25.8) of the ``normalized Huber-Mueller model'' with respect to the ``normalized distorted Huber-Mueller model'' based on method 1 (method 2), correspond to $p$-value of $7.1\times 10^{-3}$ ($8.1\times 10^{-3}$), indicating 2.7$\sigma$ (2.6$\sigma$) disfavoring of the ``normalized Huber-Mueller model'' hypothesis for \UL~spectrum. 
Similarly, the $p$-values are 0.276 (0.273) and 0.020 (0.023) for the Pu combo spectrum and the total spectrum with method 1 (method 2). 
The similar behavior in the discrimation of hypotheses between these two methods validates the feasibility of the comparison based on unfolded \nuebar~energy spectra using features of the Wiener-SVD unfolding approach.

\subsection{Discussion}
A discussion of the possible usefulness of the unfolded \nuebar~energy spectra is listed below.

1. Generation of models of the \nuebar~flux and energy spectrum for different reactor types based on the measurement at Daya Bay: 

\quad 1a) For experiments utilizing commercial LEU pressurized-water reactors~\cite{Siyeon:2017tsg,An:2015jdp,Abusleme:2020bzt}, even with fission fractions different from those of Daya Bay, the prediction can be made at 2\% precision with very little dependence on other isotopic \nuebar~flux models. 

\quad 1b) For experiments utilizing highly-enriched uranium (HEU) research reactors~\cite{Ashenfelter:2018iov,Almazan:2018wln}, the unfolded \UL~\nuebar~energy spectrum can be used directly. 

\quad 1c) For mixed-oxide (MOX) or other fuel/reactor types, which exhibit substantially different fission fractions than LEU reactor cores, some modest level of dependence on \nuebar~flux models of sub-dominant fission isotopes is expected. 

\quad 1d) For reactor-based CE$\nu$NS experiments at both HEU and LEU reactors~\cite{Aguilar-Arevalo:2019jlr, Bonet:2020awv,Agnolet:2016zir}, the CE$\nu$NS cross section can be added on the generic \nuebar~energy spectra with the IBD cross section removed. 

2. Comparison to other theoretical models in a variety of possible formats:

\quad 2a) The theoretical models can be compared with unfolded \nuebar~energy spectra directly using error matrices in the supplemental materials, at the expense of reduced precision from additional unfolding uncertainties.

\quad 2b) The comparison between theoretical models and unfolded \nuebar~energy spectra can be done with better precision using features of the Wiener-SVD unfolding method. 

\quad 2c) The comparison can also be done by getting prediction of the prompt energy spectrum with better precision using Daya Bay's detector response matrix. This procedure has a long history, and some knowledge about the detector response of Daya Bay detectors is needed.  

3. Antineutrino-based reactor monitoring applications: 

\quad 3a) The unfolded \nuebar~energy spectra provide inputs for reactor power monitoring~\cite{Vivier:2019cot} or determination of the reactor fuel types~\cite{PhysRevApplied.8.034005}. Experiments with different detector response than Daya Bay can utilize the unfolded \nuebar~energy spectra to improve their precision. 

\quad 3b) The unfolded \nuebar~energy spectra can also be used to reduce the likelihood of an undetected diversion of irradiated nuclear material~\cite{PhysRevApplied.9.014003,Stewart:2019rtd}. In addition, with inputs from isotopic \nuebar~energy spectra, the fission fractions of different isotopes for other reactors can be extracted.

\section{Summary}
\label{Sec-Summary}
Daya Bay has produced a significantly improved estimate of the \nuebar~energy spectrum from commercial nuclear reactors using 3.5 million inverse beta decay reactions. The estimate benefits from a reduction of the uncertainty in the energy response model from 1\% to 0.5\%~\cite{Adey:2019zfo} and a tripling in the IBD statistics compared to the previous results~\cite{An:2016srz}. In addition, the measurement of prompt energy spectrum for all $\bar\nu_e$ and for the $\bar\nu_e$ resulting from the main fissioning isotopes, \UL, \PuL~and \PuH, along with a new unfolding method based on the Wiener-SVD method~\cite{Tang:2017rob}, enables estimates of the respective $\bar\nu_e$ energy spectra and flux. A technique together with the relevant spectra and covariance matrices have been provided to allow a data-based prediction of the $\bar\nu_e$ flux and energy spectra for arbitrary fission fractions from a commercial reactor. The unfolded $\bar\nu_e$ energy spectra can also be directly compared with new theoretical predictions.
To avoid bias from the unfolding method and to have better precision of the comparison between theoretical models and unfolded \nuebar~energy spectra, a new method is proposed using features of the Wiener-SVD unfolding method. 
\\

\acknowledgments{
}


\vspace{10mm}

\subsection*{Appendix}
\begin{small}

\noindent{\bf Measurement of the finely-binned prompt energy spectrum of reactor antineutrinos}
\label{Sec-SpectrumAna}

The \nuebar~energy spectrum from a nuclear reactor is a sum of the \nuebar's from about 800 fission products, and it contains fine structures because of the different end-point energies (leading to sharp cutoffs) in the individual \nuebar~energy spectra from fission products with large fission yields~\cite{Sonzogni:2017voo, Dwyer:2014eka}. 
Recently an experiment with unprecedented energy resolution was proposed to search for such fine structures~\cite{Abusleme:2020bzt}. These fine structures may also be important for experiments performing precision measurements of neutrino oscillation~\cite{An:2015jdp}.
Thus, we provide a prompt energy spectrum with fine energy bins. 

To better match the energy resolution of the Daya Bay detectors, the bin width is reduced to 0.05~MeV from the previous choice of 0.25~MeV from 1 to 8~MeV. From 0.7 to 1~MeV the uncertainty is still dominated by systematic uncertainty, so only one bin is set in this energy range. 
The standard deviation of the energy resolution at Daya Bay (Eq.~\ref{Eq-EnergyResolution}) is shown in Fig.~\ref{Fig-Resolution}. The 0.05~MeV bin width is below the  standard deviation of the energy resolution at all energies. 

\begin{center}
\includegraphics[width=0.5\columnwidth]{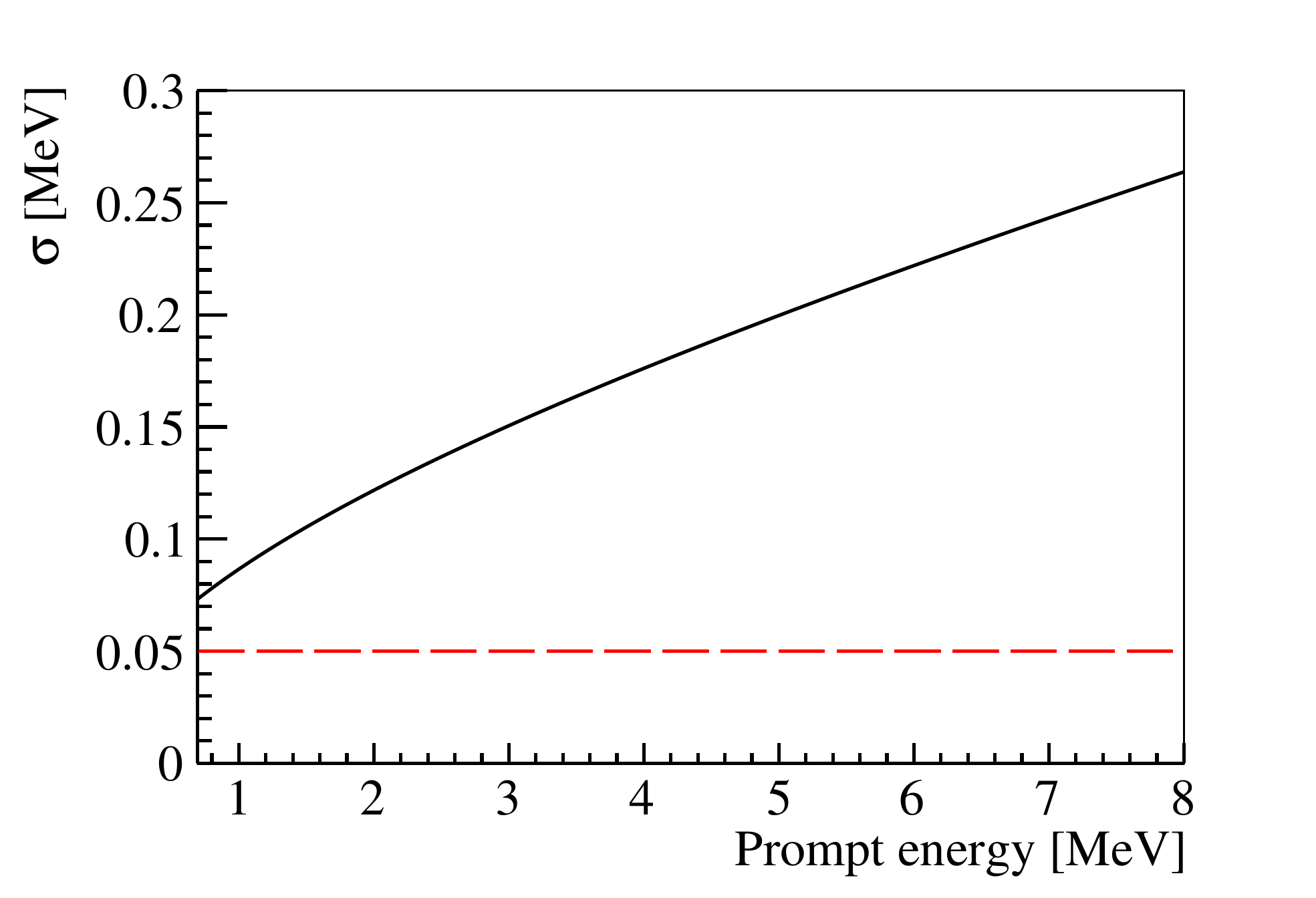}
\figcaption{\label{Fig-Resolution}  The standard deviation of the energy resolution at Daya Bay. The 0.05~MeV bin width is shown in red dash line, which is below the  standard deviation of the energy resolution at all energies. }
\end{center}

To estimate the precision of the measured prompt energy spectrum, the statistical uncertainty and systematic uncertainties are evaluated based on the understanding of the detector discussed in Sec. {\ref{Sec-DayaBayIntro}}. The systematic uncertainties contain the contribution from detector (energy nonlinearity model, energy scale difference between ADs, IAV effect) and backgrounds (accidental background, cosmogenic $^9$Li and $^8$He beta-decays, fast neutrons, Am-C neutron sources, $^{13}$C($\alpha$, n)$^{16}$O reactions, and SNF background). The backgrounds were subtracted from the total measurement of IBD samples. Here the spectral uncertainty of the measurement is focused on while the flux information ($\sim$1.5\% flux uncertainty) is not taken into account, thus the uncertainty of detection uncertainty has no impact on the spectral shape because of its energy independence. The total covariance matrix of the measured prompt energy spectrum is constructed as following:
\begin{subequations}
\renewcommand{\theequation}{A\arabic{equation}}
\begin{align}
    V = V^{\rm corr}+V^{\rm uncorr},  \label{a1}
\end{align}
where $V^{\rm corr}$ ($V^{\rm uncorr}$) represents bin-to-bin correlated (uncorrelated) uncertainty. $V^{\rm uncorr}$ contains the contribution of statistical uncertainty, and they can be evaluated by constructing covariance matrix with non-zero diagonal elements only. On the other hand, $V^{\rm corr}$ is evaluated by toy Monte Carlo as:
\begin{align}
    V_{ij}^{\rm corr}=\frac{1}{N^{\rm toy}}\sum_{i,j}^{N^{\rm toy}}(N_i^{\rm ran}-N_i^{\rm nom})(N_j^{\rm ran}-N_j^{\rm nom}),  \label{a2}
\end{align}
where $N^{\rm toy}$ is the number of toy Monte Carlo samples, $N_i^{\rm ran(nom)}$ is the random (nominal) predicted number of events at prompt energy bin $i$. In each toy Monte Carlo test the total number of events in the random predicted spectra ($\sum_i N_i^{\rm ran}$) are normalized to the nominal predicted spectrum ($\sum_i N_i^{\rm nom}$) . 
Figure~\ref{Fig-TotalUncertainty} shows the fractional size of the diagonal elements of the covariance matrix, $\sqrt{V_{ii}}/N^{\rm nom}_i$, for each component in each prompt energy bin. The elements of the correlation matrix, $V_{ij}/\sqrt{V_{ii}V_{jj}}$, for the total uncertainty is also shown in Fig.~\ref{Fig-TotalUncertainty}. The relative uncertainty ($\sqrt{V_{ii}}/N^{\rm nom}_i$) of the shape measurement is less than 1\% from 2 to 5~MeV of prompt energy. The uncertainty is dominated by the statistical uncertainty in the 2 to 8 MeV energy region. 
The measured finely-binned prompt energy spectrum and its covariance matrix are provided in the supplemental materials.

\begin{center}
\includegraphics[width=0.5\columnwidth]{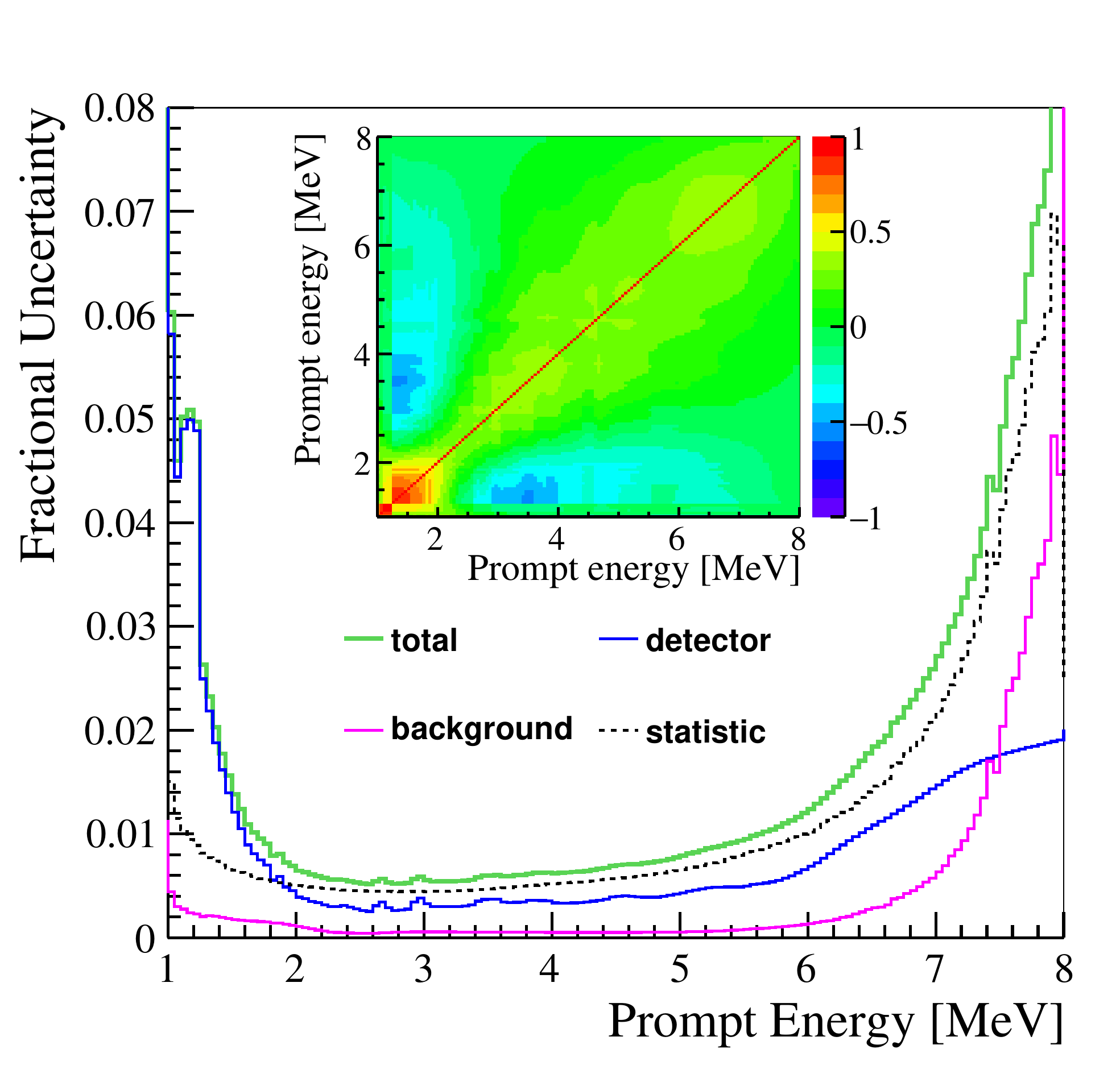}
\figcaption{\label{Fig-TotalUncertainty}  The fractional size of the diagonal elements of the covariance matrix, $\sqrt{V_{ii}}/N^{\rm pred}_i$, for each component in each prompt energy bin. Inset: the elements of the correlation matrix: $V_{ij}/\sqrt{V_{ii}V_{jj}}$. }
\end{center}

Figure~\ref{Fig-TotalMeasure} shows the finely-binned prompt energy spectrum and the comparison with models normalized to the measured number of events. Here the Huber-Mueller model \cite{Mueller:2011nm,Huber:2011wv} from the conversion method and the SM2018 model~\cite{Estienne:2019ujo} from the summation method are used for comparison. The formula of the IBD cross section from Ref.~\cite{Vogel:1999zy} is used to evaluate the detection probability of \nuebar~with different energy. The response matrix of Daya Bay detector introduced in Sec.~{\ref{Sec-DayaBayIntro}} is used to map \nuebar~energy to reconstructed prompt energy of IBD events. 
The Huber-Mueller model is provided in 0.25~MeV bins below 8~MeV of \nuebar~energy, so the content within each 0.25~MeV bin is calculated using exponential interpolation and the uncertainty of the prediction is not shown. The comparison with the measurement above 7.2~MeV of prompt energy is not shown since the content above 8~MeV of \nuebar~energy for the Huber-Mueller model is from extrapolation. 
The SM2018 model is based on the summation method and lacks an uncertainty estimate. 
Non-equilibrium effects are considered for the Huber-Mueller model based on the correction from Ref. \cite{Mueller:2011nm}, which contributed as much as $\sim$6\% additional IBD candidates for specific fissile isotopes and energy bins. The non-equilibrium effect leads to $\sim$0.71\% (with relative uncertainty of 30\%) more IBD candidates in total energy ranges. 
For the SM2018 model, the calculation with a 450-day irradiation duration ensures that the contribution from long-lived fission fragments reached equilibrium. Other information like detector DAQ livetime and detection efficiency, reactor power and working time, survival probability due to the $\theta_{13}$ driven neutrino oscillation, is all taken into account in the calculation. A spectral distortion is observed through the whole energy region for both the Huber-Mueller model and the SM2018 model, revealing that there are still things missing in current antineutrino prediction models. 

With the finely-binned spectrum more studies can be carried out. In the following, we discuss one example in detail. 

To study the fine structures and evaluate the continuity of the measurement, a method following Ref.~\cite{Sonzogni:2017voo} is used:
\begin{align}
    R_i=\frac{S_i}{S_{i+1}},  \label{a3}
\end{align}
where $S_i$ is the number of events in prompt energy bin $i$ of the measurement. 
The $R_i$ from a smooth spectrum without fine structures (e.g. Huber-Mueller model) was used to compare with the one from measurement based on the above method to evaluate the continuity. 

To avoid the comparison with a smooth model when evaluating the continuity, we propose another method based on the following formula:
\begin{align}
    R_i^{\prime}=\frac{S_i+S_{i+2}-2S_{i+1}}{S_{i+1}}.  \label{a4}
\end{align}
\end{subequations}
$R_i$ and $R_i^{\prime}$ have similar behavior of evaluating the continuity of the spectra, and $R_i^{\prime}$ is more straightforward to reveal the unsmooth structures ($R_i^{\prime}\neq0$). 
The continuity of the measurement is shown in bottom two panels of Fig. \ref{Fig-TotalMeasure}. 
The evaluated continuities of the predictions from the Huber-Mueller model and the SM2018 model are quite similar, and the $R_i^{\prime}$ of predictions is close to 0 at most of the energy region. 
To evaluate the sensitivity to fine structures at Daya Bay, the continuities of the \nuebar~energy spectra with different energy resolution are evaluated based on the SM2018 model, which are shown in Fig.~\ref{Fig-SM2018}. 
The SM2018 is based on summation method, with fine structures in the original \nuebar~energy spectrum.
In the absence of resolution effects, $R_i^{\prime}$ can reach $3\times 10^{-3}$ around 4.5~MeV on the \nuebar~energy spectra of the SM2018 model. 
After considering energy resolution of Daya Bay, $R_i^{\prime}$ decreases to $5\times 10^{-4}$. 
For experiments with $\sim$$3\%/\sqrt{E_{\rm p} [\rm MeV]}$ energy resolution~\cite{An:2015jdp,Abusleme:2020bzt}, the change on $R_i^{\prime}$ is small when considering energy resolution smearing with respect to the original \nuebar~energy spectrum. 
This observation suggests that the Daya Bay measurement is not sensitive to fine structures calculated from current nuclear databases because of the finite energy resolution. 
To confirm this conclusion, a dedicated test was done by increasing the SM2018 prediction by $\sim$10\% below 4.5~MeV to mimic the shape cut-off caused by a decay branch. While $R_i^{\prime}$ increases to 0.05 with respect to the original \nuebar~energy spectrum without energy resolution smearing, $R_i^{\prime}$ increases to $2\times 10^{-3}$ only after taking into account the energy resolution smearing. Therefore, we conclude that the measurement at Daya Bay is not sensitive to the fine structures in the original \nuebar~energy spectrum. 
Nevertheless, we report the significance of deviation from continuity as $\sigma^{\prime}_i=R_i^{\prime}/\Delta R_i^{\prime}$, where $\Delta R_i^{\prime}$ is the uncertainty of $R_i^{\prime}$. 
In addition, toy Monte Carlo tests are done by fluctuating the original prompt energy spectrum predicted by the SM2018 and Huber-Mueller models taking into account the measurement uncertainty. 
The average distribution of $\sigma^{\prime}_i$ of 10000 prompt energy spectra from toy Monte Carlo tests is consistent with the distribution of the measurement. 
The $\sigma^{\prime}_i$ distributions of the measurement and toy Monte Carlo tests are summarized in the Table~\ref{tab1}. There are two adjacent points around 7.4~MeV with larger than 2$\sigma$ deviation from continuity on the measurement. 
The distribution of $\sigma^{\prime}_i$ using measured data is consistent with the normal distribution, suggesting that the unsmooth structures on the measurement are consistent with statistical fluctuations. 
No evidence of fine structures in reactor antineutrino spectrum based on Daya Bay measurement is found because of its finite energy resolution. 
\begin{center}
\tabcaption{ \label{tab1} The significance of deviation from continuity ($\sigma^{\prime}_i$) based on the measurement and toy Monte Carlo tests. }
\footnotesize
\begin{tabular*}{80mm}{c@{\extracolsep{\fill}}ccc}
\toprule
 & $<$1$\sigma$ & 1$\sim$2$\sigma$ & 2$\sim$3$\sigma$ \\
\hline
Data & 119 & 18 & 2 \\
\hline
Toy Monte Carlo (average) & 117.6 & 20.7 & 0.7 \\
\bottomrule
\end{tabular*}
\end{center}
\end{small}

\begin{center}
\includegraphics[width=\columnwidth]{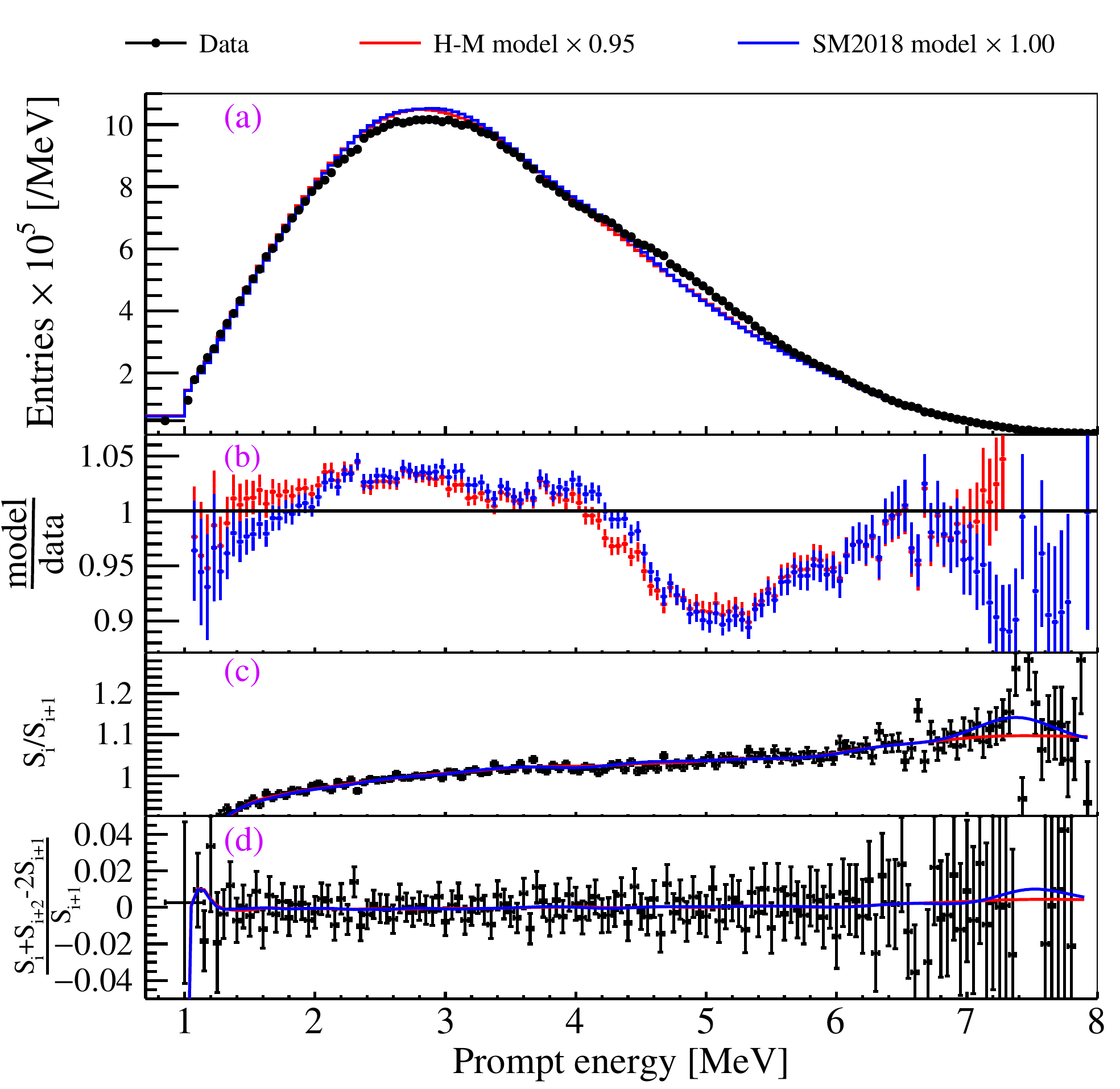}
\figcaption{\label{Fig-TotalMeasure}  (a) Measured prompt energy spectrum and the predictions from the Huber-Mueller (H-M) model and the SM2018 model (normalized to the number of measured events). The error bars on the data points ($\sim$0.6\%) represent the square root of diagonal elements of the covariance matrix for the measurement, which contains both statistical and systematic uncertainties.  
(b) Ratio of the normalized predicted spectra and the measured prompt energy spectrum. Red (Blue) histogram represents the comparison between measurement and the Huber-Mueller (SM2018) model. The error bars on the data points represent the uncertainties from measurement. 
(c) The continuity ($R_i=\frac{S_i}{S_{i+1}}$) of the measurement (black) and predictions from the Huber-Mueller model (red) and the SM2018 model (blue). 
(d) The continuity ($R_i^{\prime}=\frac{S_i+S_{i+2}-2S_{i+1}}{S_{i+1}}$) of the measurement (black) and predictions from the Huber-Mueller model (red) and the SM2018 model (blue). 
}
\end{center}

\begin{center}
\includegraphics[width=\columnwidth]{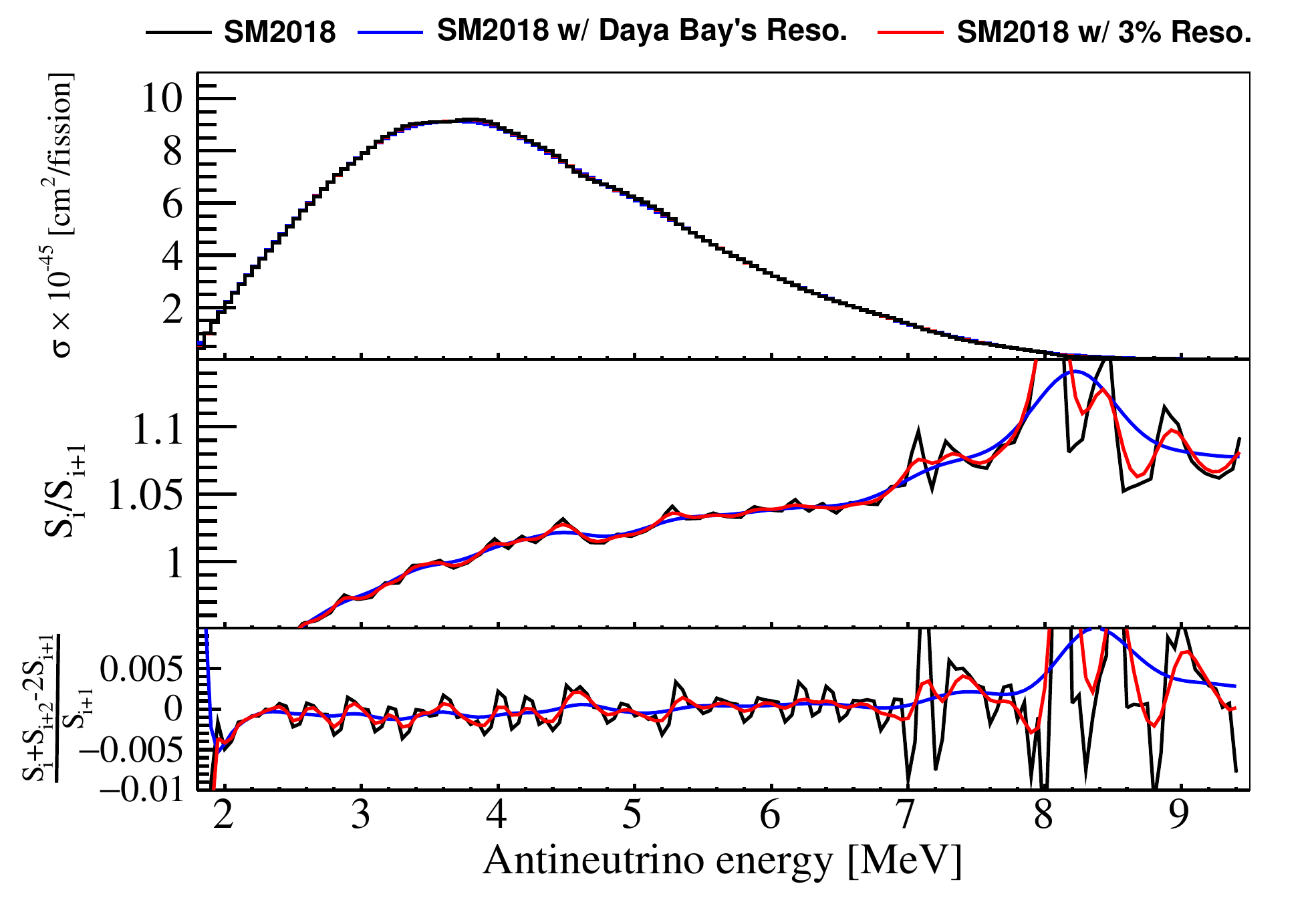}
\figcaption{\label{Fig-SM2018}  (Top panel) Original \nuebar~energy spectrum weighted by the IBD cross section based on the SM2018 model, \nuebar~energy spectrum with the energy resolution from Daya Bay, and \nuebar~energy spectrum with $3\%/\sqrt{E_{\rm p} [\rm MeV]}$ energy resolution. Here $E_{\rm p}$ is the prompt energy, while the energy resolution of \nuebar~energy $E_{\bar{\nu}_e}$ is calculated with $E_{\bar{\nu}_e}\approx E_{\rm p}+0.78~\text{MeV}$. (Middle panel) The continuity ($R_i$) of the original \nuebar~energy spectrum, the \nuebar~energy spectrum with the energy resolution from Daya Bay, and the \nuebar~energy spectrum with $3\%/\sqrt{E_{\rm p} [\rm MeV]}$ energy resolution. (Bottom panel) The continuity ($R_i^{\prime}$) of the original \nuebar~energy spectrum, the \nuebar~energy spectrum with the energy resolution from Daya Bay, and the \nuebar~energy spectrum with $3\%/\sqrt{E_{\rm p} [\rm MeV]}$ energy resolution. 
}
\end{center}


\vspace{-1mm}
\centerline{\rule{80mm}{0.1pt}}
\vspace{2mm}

\begin{multicols}{2}

\bibliographystyle{apsrev4-1}
\bibliography{Antineutrino_Energy_Spectrum_Unfolding_Based_on_the_Daya_Bay_Measurement_and_Its_Application}{}

\end{multicols}

\clearpage

\end{CJK*}
\end{document}

%% file: CPC.tex
\newcommand{\ECUST}{1}
\newcommand{\Wisconsin}{2}
\newcommand{\BNL}{3}
\newcommand{\NTU}{4}
\newcommand{\IHEP}{5}
\newcommand{\NUU}{6}
\newcommand{\TsingHua}{7}
\newcommand{\SZU}{8}
\newcommand{\ZSU}{9}
\newcommand{\NCEPU}{10}
\newcommand{\CUHK}{11}
\newcommand{\Siena}{12}
\newcommand{\UCI}{13}
\newcommand{\USTC}{14}
\newcommand{\Charles}{15}
\newcommand{\Dubna}{16}
\newcommand{\UIUC}{17}
\newcommand{\LBNL}{18}
\newcommand{\IIT}{19}
\newcommand{\BNU}{20}
\newcommand{\XJTU}{21}
\newcommand{\Yale}{22}
\newcommand{\CIAE}{23}
\newcommand{\SDU}{24}
\newcommand{\GXU}{25}
\newcommand{\VirginiaTech}{26}
\newcommand{\NCTU}{27}
\newcommand{\UC}{28}
\newcommand{\TempleUniversity}{29}
\newcommand{\DGUT}{30}
\newcommand{\UCB}{31}
\newcommand{\HKU}{32}
\newcommand{\NanKai}{33}
\newcommand{\SJTU}{34}
\newcommand{\Princeton}{35}
\newcommand{\CalTech}{36}
\newcommand{\WM}{37}
\newcommand{\NJU}{38}
\newcommand{\CGNPG}{39}
\newcommand{\NUDT}{40}
\newcommand{\IowaState}{41}
\newcommand{\CQU}{42}
\author{
 F.~P.~An(安丰鹏)$^{\ECUST}$ \and
A.~B.~Balantekin$^{\Wisconsin}$ \and
M.~Bishai$^{\BNL}$ \and
S.~Blyth$^{\NTU}$ \and
G.~F.~Cao(曹国富)$^{\IHEP}$ \and
J.~Cao(曹俊)$^{\IHEP}$ \and
J.~F.~Chang(常劲帆)$^{\IHEP}$ \and
Y.~Chang(张昀)$^{\NUU}$ \and
H.~S.~Chen(陈和生)$^{\IHEP}$ \and
S.~M.~Chen(陈少敏)$^{\TsingHua}$
% ...
\and
}

\maketitle 

\address{
\vspace{0.3cm}
{\normalsize (Daya Bay Collaboration)} \\ 
\vspace{0.3cm}
$^{\ECUST}$Institute of Modern Physics, East China University of Science and Technology, Shanghai \\ 
$^{\Wisconsin}$University~of~Wisconsin, Madison, Wisconsin 53706 \\ 
$^{\BNL}$Brookhaven~National~Laboratory, Upton, New York 11973 \\ 
$^{\NTU}$Department of Physics, National~Taiwan~University, Taipei \\ 
$^{\IHEP}$Institute~of~High~Energy~Physics, Beijing \\ 
$^{\NUU}$National~United~University, Miao-Li \\ 
$^{\TsingHua}$Department~of~Engineering~Physics, Tsinghua~University, Beijing \\ 
$^{\SZU}$Shenzhen~University, Shenzhen \\ 
$^{\ZSU}$Sun Yat-Sen (Zhongshan) University, Guangzhou \\ 
$^{\NCEPU}$North~China~Electric~Power~University, Beijing \\ 
$^{\CUHK}$Chinese~University~of~Hong~Kong, Hong~Kong \\ 
$^{\Siena}$Siena~College, Loudonville, New York  12211 \\ 
$^{\UCI}$Department of Physics and Astronomy, University of California, Irvine, California 92697 \\ 
$^{\USTC}$University~of~Science~and~Technology~of~China, Hefei \\ 
$^{\Charles}$Charles~University, Faculty~of~Mathematics~and~Physics, Prague \\ 
$^{\Dubna}$Joint~Institute~for~Nuclear~Research, Dubna, Moscow~Region \\ 
$^{\UIUC}$Department of Physics, University~of~Illinois~at~Urbana-Champaign, Urbana, Illinois 61801 \\ 
$^{\LBNL}$Lawrence~Berkeley~National~Laboratory, Berkeley, California 94720 \\ 
$^{\IIT}$Department of Physics, Illinois~Institute~of~Technology, Chicago, Illinois  60616 \\ 
$^{\BNU}$Beijing~Normal~University, Beijing \\ 
$^{\XJTU}$Department of Nuclear Science and Technology, School of Energy and Power Engineering, Xi'an Jiaotong University, Xi'an \\ 
$^{\Yale}$Wright~Laboratory and Department~of~Physics, Yale~University, New~Haven, Connecticut 06520 \\ 
$^{\CIAE}$China~Institute~of~Atomic~Energy, Beijing \\ 
$^{\SDU}$Shandong~University, Jinan \\ 
$^{\GXU}$Guangxi University, No.100 Daxue East Road, Nanning \\ 
$^{\VirginiaTech}$Center for Neutrino Physics, Virginia~Tech, Blacksburg, Virginia  24061 \\ 
$^{\NCTU}$Institute~of~Physics, National~Chiao-Tung~University, Hsinchu \\ 
$^{\UC}$Department of Physics, University~of~Cincinnati, Cincinnati, Ohio 45221 \\ 
$^{\TempleUniversity}$Department~of~Physics, College~of~Science~and~Technology, Temple~University, Philadelphia, Pennsylvania  19122 \\ 
$^{\DGUT}$Dongguan~University~of~Technology, Dongguan \\ 
$^{\UCB}$Department of Physics, University~of~California, Berkeley, California  94720 \\ 
$^{\HKU}$Department of Physics, The~University~of~Hong~Kong, Pokfulam, Hong~Kong \\ 
$^{\NanKai}$School of Physics, Nankai~University, Tianjin \\ 
$^{\SJTU}$Department of Physics and Astronomy, Shanghai Jiao Tong University, Shanghai Laboratory for Particle Physics and Cosmology, Shanghai \\ 
$^{\Princeton}$Joseph Henry Laboratories, Princeton~University, Princeton, New~Jersey 08544 \\ 
$^{\CalTech}$California~Institute~of~Technology, Pasadena, California 91125 \\ 
$^{\WM}$College~of~William~and~Mary, Williamsburg, Virginia  23187 \\ 
$^{\NJU}$Nanjing~University, Nanjing \\ 
$^{\CGNPG}$China General Nuclear Power Group, Shenzhen \\ 
$^{\NUDT}$College of Electronic Science and Engineering, National University of Defense Technology, Changsha \\ 
$^{\IowaState}$Iowa~State~University, Ames, Iowa  50011 \\ 
$^{\CQU}$Chongqing University, Chongqing \\ 
} 